\documentclass[%
twocolumn,
amsmath,
amssymb,
aps,
prl,
nofootinbib,
superscriptaddress,
floatfix,
longbibliography]{revtex4-2}

\usepackage[pdftex]{graphicx}
\usepackage{hyperref}

\usepackage{bm}
\usepackage{braket}
\usepackage{mathtools}
\usepackage{xspace}
\usepackage{array}
\usepackage{multirow}
\usepackage{comment}

\hypersetup{colorlinks = true, urlcolor = blue, linkcolor = blue, citecolor = blue}

\makeatletter

\let\MYcaption\@makecaption
\makeatother

\usepackage{subcaption}
\makeatletter
\let\@makecaption\MYcaption
\makeatother

\usepackage{color}
\usepackage[normalem]{ulem} 

\newcommand{\bk}{\bm{k}}

\begin{document}

\title{Field-induced superconductivity mediated by odd-parity multipole fluctuation}

\author{Kosuke Nogaki}
\email[]{nogaki.kosuke.83v@st.kyoto-u.ac.jp}
\affiliation{%
  Department of Physics, Kyoto University, Kyoto 606-8502, Japan
}%

\author{Youichi Yanase}
\affiliation{%
  Department of Physics, Kyoto University, Kyoto 606-8502, Japan
}%

\date{\today}

\begin{abstract}
Field-induced superconductivity has long presented a counterintuitive phenomenon and a pivotal challenge in condensed matter physics.
In this Letter, we introduce a mechanism for achieving field-induced superconductivity wherein the sublattice degree of freedom and the Coulomb interaction are tightly entwined.
Our multipole-resolved analysis elucidates that lifting the fluctuation degeneracy results in an unconventional Cooper pairing channel, thereby realizing field-induced superconductivity.
This research substantively augments the exploration of the latent potential of strongly correlated electron systems with sublattice degrees of freedom.
\end{abstract}

\maketitle

\textit{Introduction.} --- 
Superconductivity, which is typically suppressed by a magnetic field through both the Pauli and orbital-depairing effects~\cite{Tinkham2004book}, is paradoxically induced by the magnetic field in some systems. 
This counterintuitive phenomenon has garnered significant attention due to its implications for the unconventional origins of superconductivity. 
One of the well-known mechanisms of field-induced superconductivity is the Jaccarino-Peter effect~\cite{Jaccarino1962prl,Fischer1972hpa}, which states that the external magnetic field compensates for the internal field produced by magnetic ions. 
Notably, the Chevrel phase superconductor Eu$_x$Sn$_{1-x}$Mo$_6$S$_8$~\cite{Meul1984prl,Fischer1985prl} and the organic superconductors $\lambda$-(BETS)$_2$FeCl$_4$~\cite{Uji2001nature,Balicas2001prl} and $\kappa$-(BETS)$_2$FeBr$_4$~\cite{Konoike2004prb} have been associated with the Jaccarino-Peter effect. 
In addition, the decrease of Kondo scattering~\cite{Podmarkov1989sst,Podmarkov1989jetp} and the reduction of the quasi-particle renormalization effect~\cite{Tachiki1986prb} have also been discussed as other possible mechanisms.

Field-induced superconductivity in uranium-based superconductors, as observed in compounds such as UGe$_2$~\cite{Saxena2000nature,Sheikin2001prb}, URhGe~\cite{Aoki2001nature,Levy2005science}, UCoGe~\cite{Huy2007prl,Aoki2009jpsj}, and UTe$_2$~\cite{Ran2019science,Aoki2019jpsj,Ran2019naturephys,Knebel2019}, has been established experimentally. This class of phenomena has predominantly been attributed to changes in effective interactions for Cooper pairing, specifically the amplification of ferromagnetic fluctuations.
The application of a magnetic field brings these systems closer to a quantum critical point, which in turn enhances the strength of effective interactions responsible for the observed field-induced superconductivity~\cite{Hattori2012prl,Hattori2013prb,Tada2013jpcs,Hattori2014jpsj,Tokunaga2015prl,Tokunaga2016prb,Tada2016prb,Wu2017naturecom,Rosuel2023prx}.

Recently, the discovery of field-induced parity transition in CeRh$_2$As$_2$ has illuminated the role of sublattice degrees of freedom in heavy-fermion systems~\cite{Khim2021science}. Subsequent intensive experimental and theoretical works have demonstrated that local inversion symmetry breaking can enable Cooper pairs to form odd-parity pairings in the high-magnetic-field phase~\cite{Schertenleib2021prr,Mockli2021prr,Ptok2021prb,Nogaki2021prr,Mockli2021prb,Kimura2021prb,Onishi2022fem,Cavanagh2022prb,Hafner2022prx,Kibune2022prl,Kitagawa2022jpsj,Hazra2023prl,Nogaki2022prb,Landaeta2022prx,Machida2022prb,Siddiquee2022prb,Mishra2022prb,Lee2022arxiv,Semeniuk2023arxiv,Ogata2023prl,Hackner2023arxiv,Szabo2023arxiv,Christovam2023arxiv,Szabo2023arxiv2,Chen2023arxiv,Wu2023arxiv,Ishizuka2023arxiv}. Interestingly, similar sublattice structures in the unit cells are also inherent in the uranium compounds mentioned earlier. In addition, field-induced superconductivity has been reported in a locally-noncentrosymmetric cerium-based superconductor CeSb$_2$~\cite{Canfield1991apl,Squire2023prl} and in magic-angle twisted trilayer graphene~\cite{Park2021nature,Cao2021nature}. Both uranium- and cerium-based superconductors, as well as moiré systems, may have superconducting states driven by electron correlation effects. Given this, theoretical studies focusing on the strong correlation effect, sublattice degrees of freedom, and the magnetic field, are of significant interest. 
However, most of the previous theoretical studies have been based on the weak coupling theory.
In particular, it should be noted that previous research has often assumed \textit{degenerate} interactions in the sublattice degrees of freedom.

\textit{Effective Action.} --- 
In this Letter, we introduce a mechanism for field-induced superconductivity that originates from \textit{degeneracy-lifted} pairing interactions in sublattice degrees of freedom.
A theoretical basis for strongly correlated superconductors is the following effective action,
\begin{align}
    \mathcal{S}_{\mathrm{eff}}[\bar{\psi},\psi] 
    &=\mathcal{S}_{\mathrm{eff, \, 0}}[\bar{\psi},\psi]  
    +\mathcal{S}_{\mathrm{eff, \, int}}[\bar{\psi},\psi] \notag \\
    &=\sum_k\bar{\psi}_{k,\alpha} (-i\omega_n \delta^{\alpha\beta}+\mathcal{H}^{\alpha\beta}_{\bk}+\Sigma^{\alpha\beta}_k) \psi_{k,\beta} \notag \\
    &+ \sum_{k,k',q} \bar{\psi}_{k+q,\alpha} \psi_{k,\beta} \Gamma^{\alpha\beta\gamma\delta}_{q} \bar{\psi}_{-k',\delta} \psi_{-k'+q,\gamma},
\end{align}
where abbreviated notation of $k=(\bm{k},i\omega_n)$, $q=(\bm{q},i\nu_n)$, and $\alpha=(s,\sigma)$ are employed.
Here, the momentum $\bm{k}$ and the fermionic and bosonic Matsubara frequencies $\omega_n=(2n+1)\pi T$, $\nu_n=2n\pi T$ stand for the space-time dependence of the electron field $\psi_{k,\alpha}$ and the correlation functions.
The index $s$ and $\sigma$ represent the spin and sublattice degrees of freedom, respectively.
The $\mathcal{H}^{\alpha\beta}_{\bm{k}}$ is the single-particle Hamiltonian.
We introduce the self-energy ($\Sigma$) and the vertex function ($\Gamma$); the former causes the renormalization of mass and damping of quasi-particles, while the latter drives the system toward superconductivity. 
These quantities obey the Ward-Takahashi identity: $\Sigma^{\alpha\beta}_k = \sum_{q}\Gamma^{\alpha\gamma\beta\delta}_q G^{\delta\gamma}_{k-q}$~\cite{Ward1950pr,Takahashi1957inc}.
Here, $G^{\alpha\beta}_k=-\langle\psi_{k,\alpha}\bar{\psi}_{k,\beta}\rangle$ describes the single-particle Green function.
While the $k$ and $k'$ dependence of the vertex function are ignored for brevity, the extension of the following discussion to include such momentum dependence is straightforward~\cite{Kozii2015prl,Sumita2020prr}.

By virtue of diagrammatic techniques (e.g. the fluctuation exchange (FLEX) approximation~\cite{Bickers1989prl,Bickers1989ap,Bickers1991prb} or Parquet approximation~\cite{Roulet1969pr,Bickers2004book}), the above effective action can be derived from a bare action:
\begin{align}
    \mathcal{S}_{\mathrm{bare}}[\bar{\psi},\psi] &= 
    \sum_k\bar{\psi}_{k,\alpha} (-i\omega_n \delta^{\alpha\beta}+\mathcal{H}^{\alpha\beta}_{\bk}) \psi_{k,\beta} \notag \\
    &+ \sum_{k,k',q} \bar{\psi}_{k+q,\alpha} \psi_{k,\beta} \Gamma^{0,\alpha\beta\gamma\delta}_{q} \bar{\psi}_{-k',\delta} \psi_{-k'+q,\gamma}.
\end{align}
In this study, we focus on two-sublattice superconductors, including bilayer superconductors and two-fold non-symmorphic crystalline superconductors.
In the bare action, $\mathcal{H}^{\alpha\beta}_{\bk}$ represents Hamiltonian of the two-sublattice system: $\mathcal{H}_{\bk} = \varepsilon_{\bk}s_0\otimes\sigma_0 + \alpha \bm{g}_{\bk}\cdot\bm{s}\otimes\sigma_z +t_{\perp}s_0\otimes\sigma_x - \mu_{\mathrm{B}}H s_z\otimes\sigma_0$.
Here, $\varepsilon_{\bk}=-2t(\cos k_x+\cos k_y)+4t'\cos k_x\cos k_y-\mu$ and $t_{\perp}$ represent intra- and inter-sublattice hopping term, respectively, and sublattice-dependent $\bm{g}_{\bk}\cdot\bm{s}$ term represents staggered Rashba-type spin-orbit coupling which are originated from local inversion symmetry breaking~\cite{Yanase2010jpsj,Fischer2011prb,Maruyama2012jpsj,Yoshida2012prb,Maruyama2013jpsj,Yoshida2013jpsj,Yoshida2014jpsj,Yoshida2015prl,Hitomi2016jpsj,Mockli2018prb,Schertenleib2021prr,Lu2021prb,Skurativska2021prr}.
The g-vector $\bm{g}_{\bk} = [-\partial \varepsilon_{\bm{k}}/\partial k_y ,\partial \varepsilon_{\bm{k}}/\partial k_x,0]$ introduces the momentum-dependent spin polarization~\cite{Bauer2012book}.
The $H$ represents the Zeeman magnetic field parallel to the $z$-axis. 
The interaction term in the bare action is the Hubbard-type on-site Coulomb repulsion:
$\mathcal{S}_{\mathrm{bare, \, int}} = U\sum_{\sigma}\bar{\psi}_{i,\uparrow,\sigma}\psi_{i,\uparrow,\sigma}\bar{\psi}_{i,\downarrow,\sigma}\psi_{i,\downarrow,\sigma}$.
The bare interaction tensor $\Gamma^0$ is obtained from the above Hubbard interaction~\cite{supplement}.

The internal degrees of freedom of the two-sublattice model are classified by the augmented multipole operator $\hat{\mathcal{Q}}$~\cite{Watanabe2018prb,Hayami2018prb,Yatsushiro2021prb}: $\hat{\mathcal{Q}}^{\mu\nu} = \sum_{k}\bar{\psi}_{k+q,\alpha} \mathcal{Q}^{\mu\nu}_{\alpha\beta} \psi_{k,\beta}$,
where $\mathcal{Q}^{\mu\nu}=\bar{s}^{\mu}\otimes\bar{\sigma}^{\nu}$ satisfies the normalization condition $\mathrm{tr}[\mathcal{Q}\mathcal{Q}^{\dagger}]=1$~\cite{supplement}.
Here, $\bar{s}^{\mu}=s^{\mu}/\sqrt{2}$ ($\bar{\sigma}^{\mu}=\sigma^{\mu}/\sqrt{2}$) are the normalized Pauli and unit matrices.
The interaction term $\mathcal{S}_{\mathrm{eff, \, int}}$ can be expressed as the sum of the bi-linear interaction of the multipoles~\cite{Sugano1970textbook,Simmonett2014jcp,Coury2016prb,Bunemann2017jpcm,Iimura2021prb}:
\begin{equation}
    \mathcal{S}_{\mathrm{eff, \, int}}[\bar{\psi},\psi] \approx 
    \sum_{\mathcal{Q},q} \hat{\mathcal{Q}}_q V^{\mathcal{Q}}_q \hat{\mathcal{Q}}_{-q},
\end{equation}
where $V^{\mathcal{Q}}_q =\mathcal{Q}_{\alpha\beta}\Gamma^{\beta\alpha\gamma\delta}_{q}\mathcal{Q}_{\gamma\delta}$ describes the coupling constants of interaction between the augmented multipoles.
For the sake of brevity, we here omit interactions between different multipoles.

From the multipole resolved interaction, a zero-momentum ($q=k-k'=0$) Cooper pairing interaction can be obtained~\cite{Kozii2015prl,Sumita2020prr}.
The intra-sublattice even- and odd-parity multipole fluctuations, denoted by $\bar{\sigma}^0$ and $\bar{\sigma}^z$, result in the following Cooper pairing interactions~\cite{Sumita2020prr}:
\begin{align}
    \mathcal{S}_{\sigma_0}[\bar{\psi},\psi] = 
    \frac{1}{2}\sum_{k,k'}V^{\sigma^0}_{k-k'}\{
    &\hat{\mathcal{P}}^{0, \dagger}_{k}  \hat{\mathcal{P}}^{0}_{k'}
    +\hat{\mathcal{P}}^{z, \dagger}_{k}  \hat{\mathcal{P}}^{z}_{k'} \notag \\
    +&\hat{\mathcal{P}}^{x, \dagger}_{k}  \hat{\mathcal{P}}^{x}_{k'}
    +\hat{\mathcal{P}}^{y, \dagger}_{k}  \hat{\mathcal{P}}^{y}_{k'}
    \},
    \label{eq:sigma0}
\end{align}
\begin{align}
    \mathcal{S}_{\sigma_z}[\bar{\psi},\psi] = 
    \frac{1}{2}\sum_{k,k'}V^{\sigma^z}_{k-k'}\{
    &\hat{\mathcal{P}}^{0, \dagger}_{k}  \hat{\mathcal{P}}^{0}_{k'}
    +\hat{\mathcal{P}}^{z, \dagger}_{k}  \hat{\mathcal{P}}^{z}_{k'} \notag \\
    -&\hat{\mathcal{P}}^{x, \dagger}_{k}  \hat{\mathcal{P}}^{x}_{k'}
    -\hat{\mathcal{P}}^{y, \dagger}_{k}  \hat{\mathcal{P}}^{y}_{k'}
    \},
    \label{eq:sigmaz}
\end{align}
where $\hat{\mathcal{P}}^{\mu} = \psi_{\alpha}\bar{\sigma}^{\mu}_{\alpha\beta} \psi_{\beta}$.
In the degenerate case, where even- and odd-parity multipole interactions have the same coupling constant (i.e., $V_{k-k'}:= V^{\sigma^0}_{k-k'}=V^{\sigma^z}_{k-k'}$), we obtain:
\begin{align}
    \mathcal{S}_{\mathrm{degenerate}} = 
    \sum_{k,k'}V_{k-k'}\{
    \hat{\mathcal{P}}^{0, \dagger}_{k}  \hat{\mathcal{P}}^{0}_{k'}
    +\hat{\mathcal{P}}^{z, \dagger}_{k}  \hat{\mathcal{P}}^{z}_{k'} \}.
\end{align}
This is nothing but the frequently assumed pairing interaction for two-sublattice models~\cite{Fischer2011prb,Youn2012prb,Goryo2012prb,Maruyama2012jpsj,Yoshida2012prb,Maruyama2013jpsj,Yoshida2013jpsj,Yoshida2014jpsj,Mockli2018prb,Schertenleib2021prr}.
When the degeneracy is lifted, the second-line terms in Eqs.~(\ref{eq:sigma0}) and  (\ref{eq:sigmaz}) could result in unconventional inter-sublattice Cooper pairing channel.
Field-induced superconductivity, a main result of this Letter, is attributed to such degeneracy-lifted interactions, which are ubiquitous in strongly correlated systems.
The FLEX approximation extended to spin-orbit-coupled two-sublattice systems is employed to derive the effective action in this work~\cite{Luttinger1960pr,Luttinger1960pr2,Baym1961pr,Baym1962pr}.
Hereafter, we set $t'=0.3$, $\mu_{\mathrm{B}}=1$, and $U=3.9$ with a unit of energy $t=1$ and determine the chemical potential so that the electron density per site $n$ is $0.85$. 
In the numerical study, we use $64\times64$ $\bm{k}$-meshes, and $16384$, $8192$, or $4096$ Matsubara frequencies for $T=0.004$, $0.004<T<0.01$, or $0.01\leq T$, respectively~\cite{supplement}.

\textit{Odd-parity multipole fluctuation.} --- 
For the later analysis, we discuss here the multipole susceptibilities without the self-energy, as detailed in the Supplemental materials~\cite{supplement}:
\begin{align}
\chi^{\mathcal{Q}}(q) &\approx\frac{\chi^{0,\mathcal{Q}}(q)}{1-U^{\mathcal{Q}}\chi^{0,\mathcal{Q}}(q)}, \\
\chi^{0,\mathcal{Q}}(q) &=\sum_{\bk}\mathcal{Q}^{\bk-\bm{q},\bk}_{\eta\zeta} \mathcal{Q}^{\bk,\bk-\bm{q}}_{\zeta\eta} L_{\zeta\eta}(\bm{k},\bm{q},i\nu_n),
\end{align}
where $\mathcal{Q}^{\bk,\bk'}_{\zeta\eta} = \braket{u_{\zeta,\bk}|\mathcal{Q}|u_{\eta,\bk'}}$ and $L_{\zeta\eta}(\bm{k},\bm{q},i\nu_n)=-\frac{1}{N}\{f(\varepsilon_{\eta,\bm{k}-\bm{q}})-f(\varepsilon_{\zeta,\bm{k}})\}/\{i\nu_n + \varepsilon_{\eta,\bm{k}-\bm{q}} -\varepsilon_{\zeta,\bm{k}}\}$ represent the matrix element of the multipole operator $\mathcal{Q}$ and the momentum-resolved Lindhard function between bands $\zeta$ and $\eta$, respectively. The eigenvector and eigenvalue of a band $\zeta$ of the Hamiltonian are denoted by $\ket{u_{\zeta,\bk}}$ and $\varepsilon_{\zeta,\bk}$. The symbol $N$ represents the number of unit cells, and $f(\varepsilon)$ stands for the Fermi-Dirac distribution function.
To activate the multipole $\mathcal{Q}$ fluctuation, a sizable matrix element $\mathcal{Q}^{\bk-\bm{q},\bk}_{\eta\zeta}$ and a positive interaction for the multipole, characterized by $U^{\mathcal{Q}}>0$, are required. Thus, the electronic structure of the system inherently dictates the enhanced multipole fluctuation.

\begin{figure}[tbp]
 \begin{center}
\includegraphics[keepaspectratio, scale=0.18]{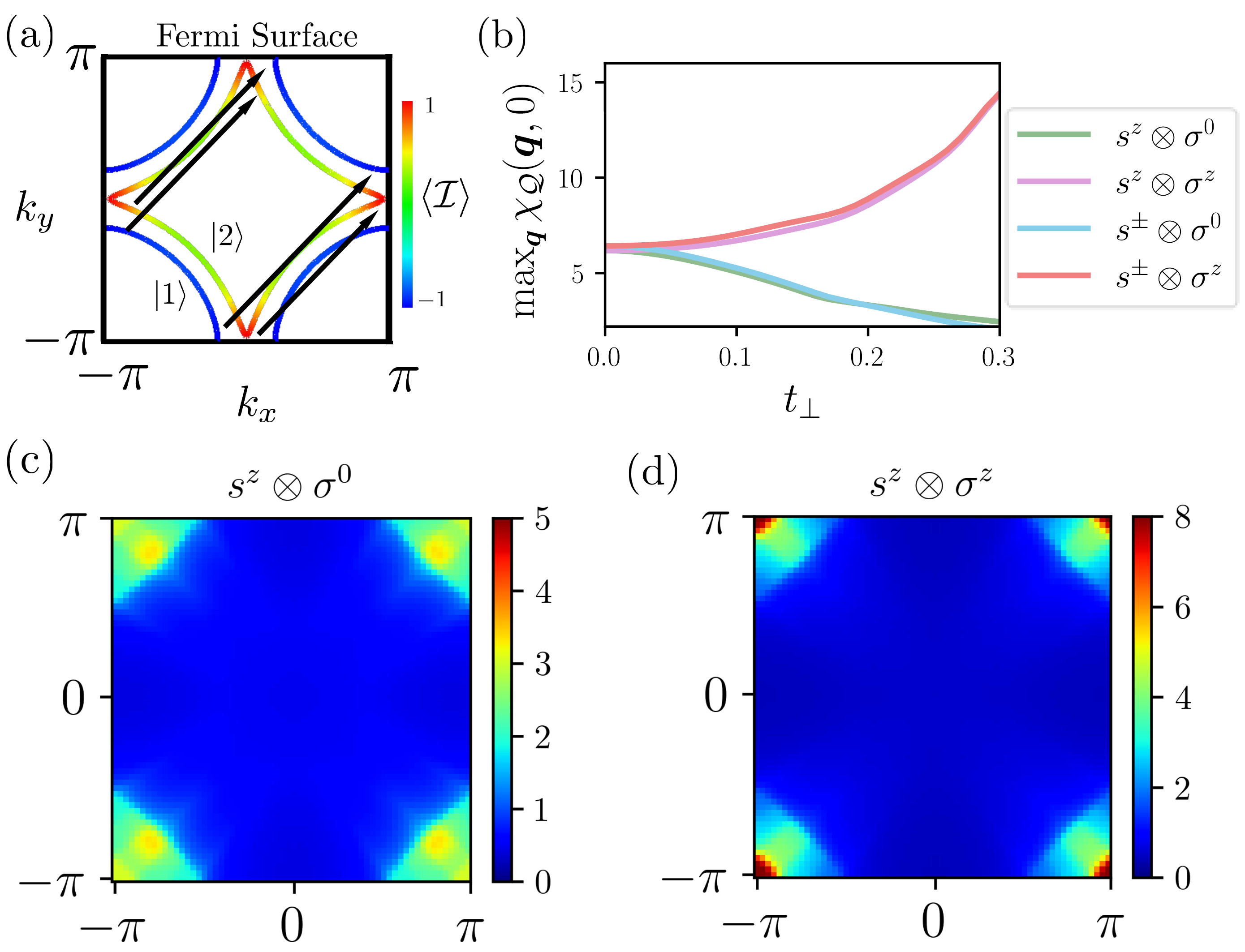}
  \end{center}
  \caption{
  (a) The Fermi surface of the two-sublattice tight-binding model with the model parameters: $\alpha=0.2,\,t_{\perp}=0.2$.
  The coloration on the Fermi surface signifies the expectation value of the inversion symmetry operator $\mathcal{I}=s_0\otimes\sigma_z$.
  (b) The inter-sublattice hopping $t_{\perp}$ dependence of static multipole fluctuations. 
  The maximum of the even-parity (odd-parity) longitudinal (transverse) magnetic multipole susceptibilities are shown. 
  Other multipole fluctuations are negligibly small.
  We assume $\alpha=0.2$, $t_{\perp}=0.2$ and $T=0.01$.
  (c), (d) The momentum dependence of the even-parity 
  and odd-parity longitudinal magnetic susceptibilities, respectively.
  }
  \label{fig:suscep}
\end{figure}

Herein, we demonstrate that a two-sublattice structure intrinsically favors odd-parity multipole fluctuations. Figure~\ref{fig:suscep}(a) illustrates the Fermi surfaces of our two-sublattice tight-binding model with two bands, labeled $\ket{1}$ and $\ket{2}$. 
Our system exhibits a type-II van Hove singularity resulting from the Rashba-type spin-orbit coupling~\cite{Yao2015prb,Wu2019prb,Nogaki2020prb}. Specifically, the type-II van Hove singularities are located around $\bm{k}=(\pm \delta,\pi)$, $(0,\pi \pm \delta)$, $(\pi,\pm \delta)$, and $(\pi \pm \delta,0)$, away from the time-reversal invariant momentum that satisfies $\bm{K} = -\bm{K}$ modulo reciprocal lattice vectors. 
Figure~\ref{fig:suscep}(a) displays the expectation values of the inversion symmetry operator, $\mathcal{I}=s_0\otimes\sigma_z$, on the Fermi surfaces.
The bonding and anti-bonding orbitals defined as $\ket{\mathrm{BO}} \equiv (1,1)_{\sigma}^\top/\sqrt{2}$ and $\ket{\mathrm{ABO}} \equiv (1,-1)_{\sigma}^\top/\sqrt{2}$ satisfy $\mathcal{I}\ket{\mathrm{BO}}=\ket{\mathrm{BO}}$ and $\mathcal{I}\ket{\mathrm{ABO}}=-\ket{\mathrm{ABO}}$.
Considering that the expectation values of $\mathcal{I}$ on the Fermi surfaces are close to $\pm 1$, we find that the wave functions around the type-II van Hove singularities are well approximated by either the bonding or anti-bonding orbitals.

In itinerant electron systems, multipole fluctuations emerge from the nesting of Fermi surfaces especially around van Hove singularities~\cite{Moriya2000ap,Yanase2003pr}. Two potential nesting scenarios exist: one involves nesting within the same Fermi surface, either bonding to bonding or anti-bonding to anti-bonding. The other involves nesting between different Fermi surfaces, bonding to anti-bonding. The nesting vectors corresponding to each of these scenarios are illustrated in Fig.~\ref{fig:suscep}(a). Notably, while the nesting vectors connecting the same Fermi surface are not equivalent to each other, those connecting different Fermi surfaces are equivalent.
As a result, in the two-sublattice model, nesting of different Fermi surfaces strongly enhances multipole fluctuations with operators having pronounced matrix elements between the Fermi surfaces $\ket{1}$ and $\ket{2}$.

Employing the previously approximated wave functions, we can roughly evaluate the matrix element of the sublattice operator: 
$\braket{1|\sigma^0|1} \approx \braket{1|\sigma^z|2} \approx 1$ and $\braket{1|\sigma^0|2} \approx \braket{1|\sigma^z|1} \approx 0$.
From this, it becomes evident that nesting within the same Fermi surface enhances even-parity multipole fluctuations while nesting between different Fermi surfaces results in odd-parity fluctuations.
Combining these analyses, we conclude that the two-sublattice structure favors odd-parity multipole fluctuations, especially for large $t_{\perp}$ and chemical potentials near the van Hove singularity.

The Hubbard-type Coulomb interaction can be expressed in the multipole basis as:
\begin{equation}
 \mathcal{S}_{\mathrm{int}} = -U\sum_{\nu=0,z}\hat{\mathcal{Q}}^{0\nu}_{q}\hat{\mathcal{Q}}^{0\nu}_{-q}
    +U\sum_{\substack{\mu=x,y,z\\ \nu=0,z}}\hat{\mathcal{Q}}^{\mu\nu}_{q}\hat{\mathcal{Q}}^{\mu\nu}_{-q}.
    \label{eq:bare_hubbard}
\end{equation}
This expression for the multipole-resolved interaction implies that the Coulomb interaction equally enhances the even-parity and odd-parity magnetic multipoles. Figure~\ref{fig:suscep}(b) depicts the dependence of these multipole fluctuations on $t_{\perp}$, calculated using the FLEX approximation. As the inter-sublattice hopping parameter, $t_{\perp}$, increases, odd-parity fluctuations are notably enhanced while even-parity fluctuations are suppressed, consistent with the above discussions.

Figures~\ref{fig:suscep}(c)-(d) show the momentum dependence of multipole susceptibilities. The even-parity longitudinal magnetic fluctuation represented by the multipole operator $\mathcal{Q}^{z0}= \bar{s}^z\otimes\bar{\sigma}^0$ exhibits a double peak structure around $\bm{Q}\sim(\pi,\pi)$ and $(\pi-\delta,\pi-\delta)$ [Fig.~\ref{fig:suscep}(c)]. In contrast, the odd-parity longitudinal magnetic fluctuation given by the multipole operator $\mathcal{Q}^{zz}=\bar{s}^z\otimes\bar{\sigma}^z$ presents a single peak structure around $\bm{Q}\sim(\pi,\pi)$ [Fig.~\ref{fig:suscep}(d)]. These findings further substantiate the previous analysis of the nesting and wave functions of the Fermi surfaces, offering a more quantitative view. 
Note that the transverse magnetic fluctuations manifest a similar momentum dependence to the longitudinal ones.

It's worth noting that the Hubbard-type Coulomb interaction, as expressed by Eq.~(\ref{eq:bare_hubbard}), has a degenerate form when viewed on a multipole basis. 
This suggests that all magnetic multipole interactions are equivalent at the mean-field level of analysis.
However, the many-body effects lift the degeneracy through the intrinsic properties of the wave function of itinerant electrons, leading to the dominant odd-parity multipole fluctuation at low energies.

\begin{figure}[tbp]
 \begin{center}
    \includegraphics[keepaspectratio, scale=0.43]{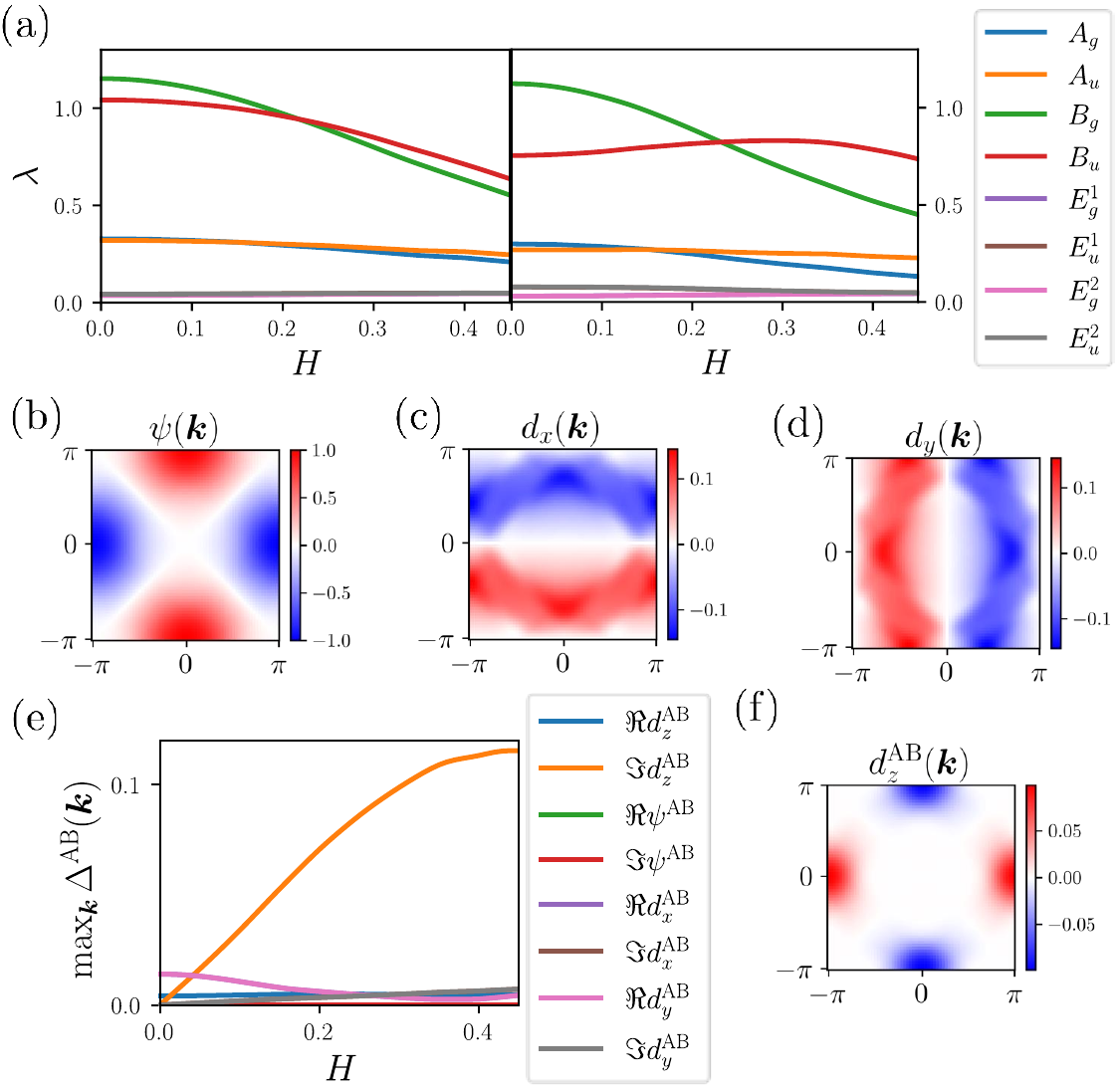}
  \end{center}
  \caption{ 
  (a) The magnetic field dependence of eigenvalues of the \'{E}liashberg equation for each irreducible representation. 
  We assume $\alpha=0.2$ and $T=0.01$.
  (Left) $t_{\perp}=0.1$. 
  (Right) $t_{\perp}=0.2$.
  Superconducting instabilities are classified by the irreducible representation of the point group of the system, $C_{4h}$.
  The superscript of $E_{g/u}^{1,2}$ representations expresses the degeneracy lifted by time-reversal symmetry breaking due to the magnetic field.
  (b-d) The momentum dependence of intra-sublattice spin-singlet and spin-triplet gap functions, $\psi(\bk)$ and $\bm{d}(\bk)$, of the $B_u$ representation for $H=0.15$. 
  Results for the $B_g$ representation are almost the same as the figures.
  (e) The magnetic field dependence of the component-resolved weight of the inter-sublattice gap function.
  (f) The momentum dependence of the inter-sublattice spin-triplet gap function, $\Im d^{\mathrm{AB}}_z(\bk)$.
  }
  \label{fig:delta}
\end{figure}

\begin{figure*}[tbp]
 \begin{center}
\includegraphics[keepaspectratio, scale=0.5]{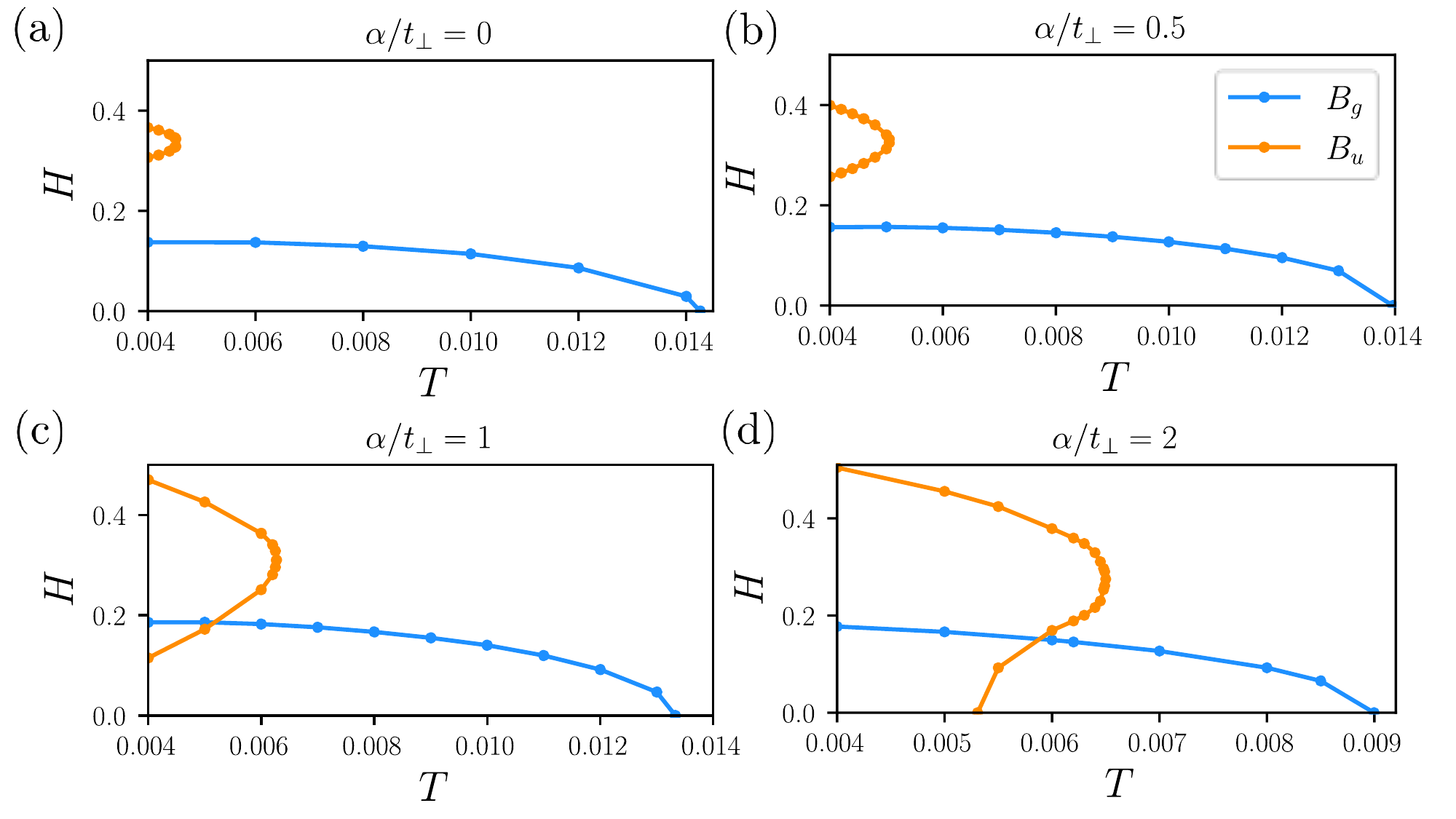}
  \end{center}
  \caption{(a)-(d) $H$-$T$ phase diagrams of the two-sublattice Rashba-Hubbard model for $\alpha/t_{\perp}=0$, $0.5$, $1$, $2$.
  We show the superconducting transition lines of the even-parity $B_{g}$ and odd-parity $B_{u}$ states, on which eigenvalues of the \'{E}liashberg equation become unity.
  }
  \label{fig:phases}
\end{figure*}

\textit{Superconductivity.} --- 
We here study superconductivity predominantly mediated by the odd-parity multipole fluctuations using the linearized \'{E}liashberg equation~\cite{supplement}. Figure~\ref{fig:delta}(a) depicts the magnetic field dependence of the eigenvalues of this equation. In the left figure for $t_{\perp}=0.1$, the typical behavior of superconductivity under an external magnetic field is evident. All eigenvalues across all irreducible representations are suppressed by the magnetic field. Notably, at $H=0.22$, the eigenvalue curves of the $B_g$ and $B_u$ representations intersect. This intersection signals a phase transition from even-parity to odd-parity superconductivity, reminiscent of phenomena observed in CeRh$_2$As$_2$~\cite{Khim2021science,Yoshida2012prb,Nogaki2022prb}. In contrast, at $t_{\perp}=0.2$ [right figure of Fig.~\ref{fig:delta}(a)], the eigenvalue for the $B_{u}$ representation increases upon magnetic field application, while that of $B_{g}$ diminishes. Such behavior suggests the possible emergence of field-induced odd-parity superconductivity in the two-sublattice strongly correlated electron systems.

We delve into the mechanism behind this field-induced superconductivity. Initially, the intra-sublattice pair potential, given by $\Delta^{\mathrm{intra}}_{B_u}(k) = \psi(\bk) is_y\otimes\sigma_z + \bm{d}(\bk)\cdot\bm{s} \, is_y \otimes\sigma_0$, is illustrated in Figs.~\ref{fig:delta}(b)-(d). Influenced by the anti-ferromagnetic fluctuation, the spin-singlet component shows a $d_{x^2-y^2}$-wave form. In addition, the spin-triplet components induced by the spin-orbit coupling display a $p$-wave momentum dependence. 
Interestingly, these gap functions are relatively impervious to the external magnetic field~\cite{Khim2021science,Yoshida2012prb,Nogaki2022prb}.
Subsequently, the inter-sublattice pair potentials are analyzed. In Fig.~\ref{fig:delta}(e), the component-resolved magnetic field dependence of inter-sublattice gap functions is shown. Evidently, the magnetic field induces a sizable spin-triplet and inter-sublattice anti-symmetric pair potential, 
\begin{equation}
\Im d_z^{\mathrm{AB}}(\bk) s_z i s_y \otimes \sigma_y,
\label{eq:inter-sublattice_pair}
\end{equation}
which is prohibited at the zero magnetic field by time-reversal symmetry. The momentum dependence of this pair potential is depicted in Fig.~\ref{fig:delta}(f). Notably, Eq.~(\ref{eq:inter-sublattice_pair}) represents a $\sigma_y$-component in the sublattice degree of freedom. This implies that the pairing channel of $\mathcal{P}^{y}_{k}$ in Eqs.~(\ref{eq:sigma0}) and (\ref{eq:sigmaz}), which arises from the lifting of degeneracy between even-parity and odd-parity multipole fluctuations, plays a pivotal role in the manifestation of field-induced superconductivity.
In essence, the Cooper pairing inherent in Eq.~(\ref{eq:inter-sublattice_pair}) is fundamentally rooted in the multipole-mediated interactions, Eqs.~(\ref{eq:sigma0}) and (\ref{eq:sigmaz}), and the disruption of the time-reversal symmetry by the external magnetic field allows this pairing to emerge. As a result, the field-induced superconductivity occurs in the two-sublattice system through a cooperative interplay between the odd-parity multipole fluctuation and the magnetic field.

Further support for our interpretation of the mechanism behind field-induced superconductivity comes from considerations based on Feynman diagram analyses~\cite{supplement}. The gap function outlined in Eq.~(\ref{eq:inter-sublattice_pair}) plays a crucial role in facilitating the coupling between intra-sublattice gap functions through a second-order scattering process.

\textit{Phase diagram.} --- 
Figures~\ref{fig:phases}(a)-(d) show the phase diagrams for $\alpha/t_{\perp}=0$, $0.5$, $1$, and $2$ with $t_{\perp}=0.2$.
As expected, the odd-parity superconducting state exhibits field-induced behaviors in all cases. It is noteworthy that even in the case of $\alpha=0$ (i.e., without spin-orbit coupling) the external magnetic field induces the odd-parity superconducting phase. Thus, the field-induced superconductivity does not require spin-orbit coupling. However, larger spin-orbit coupling renders the field-induced odd-parity superconducting phase more stable, as the transition temperature increases. Indeed, with spin-orbit coupling, the gap function in Eq.~(\ref{eq:inter-sublattice_pair}) incorporates intra-band components:
\begin{equation}
\Delta^{\pm}(\bm{k}) = \frac{d_z^{\rm AB}(\bm{k})}{|\bm{g}(\bm{k})|^2+t^2_{\perp}}\left\{\mp\tilde{\bm{d}}\cdot\tilde{\bm{s}}+i\tilde{\psi}(\bm{k})\tilde{s_0}\right\}i\tilde{s}_y, 
\end{equation}
where $\tilde{\psi}(\bk) = |\bm{g}(\bm{k})|^2$ and $\tilde{\bm{d}}=[g_y(\bm{k}),g_x(\bm{k}),0]$. Consequently, the spin-orbit coupling significantly enhances the thermodynamic stability of the odd-parity superconducting phase.

The field-induced odd-parity superconducting state suffers from the Pauli depairing effect of Cooper pairs at extremely high fields, $H>0.3$. 
Thus, the observed nonmonotonic behavior of the phase transition line of $B_u$ state can result from the competition between the field-enhancement mechanism, associated with the inter-sublattice gap function, and the Pauli depairing effect.

\textit{Summary.} --- 
In summary, we have conducted a multipole-resolved analysis for unconventional superconductivity in strongly-correlated two-sublattice systems. The lifting of degeneracy between even- and odd-parity multipole fluctuations gives rise to the unconventional pairing channel. We demonstrated that the two-sublattice structure inherently favors multipole fluctuations with a predominating odd-parity nature, which induce sublattice-antisymmetric pairing only when the magnetic field is applied. Consequently, the field-induced superconductivity occurs. Notably, the obtained phase diagram reveals the field-reentrant odd-parity superconducting states.

Field-induced superconductivity within the bilayer model has also been proposed in previous studies~\cite{Houzet2002el,Buzdin2005prl,Tollis2006prb,Buzdin2007pcs,Montiel2011prb}. In these theories, a magnetic field is posited to shift the energy levels of electronic states, thereby facilitating unconventional inter-band Cooper pairing. However, these models assume isotropic interaction within the layers, which distinctly sets them apart from the mechanism we present in this Letter. Note that the cornerstone of our proposal is the anisotropic effective interaction, a direct result of the degeneracy-lifted multipole fluctuation.

Finally, we would like to highlight a possible application of our theory. The electronic structure of magic-angle twisted trilayer graphene consists of a flat band from the moiré structure and a dispersive Dirac band in the absence of a displacement field~\cite{Khalaf2019prb,Carr2020nanoletter,Chao2021prb}. The flat band potentially enhances the degenerated multipole fluctuation which is ensured by symmetry. For instance, fifteen-fold degenerate fluctuations protected by SU(4) symmetry have been proposed in magic-angle twisted bilayer graphene~\cite{Onari2022prl}. The introduction of a displacement field leads to the hybridization of these bands, which could result in the lifting of the degeneracy in multipole fluctuations~\cite{Cao2021nature}.
Application of an external magnetic field possibly induces unconventional Cooper pairing through the degeneracy-lifted multipole fluctuations as introduced in this Letter. This mechanism might explain the magnetic field-reentrant superconductivity observed in magic-angle twisted trilayer graphene~\cite{Cao2021nature}. Other potential candidates for application of this theory include uni-axially strained CeRh$_2$As$_2$ and pressurized CeSb$_2$~\cite{Squire2023prl}. The pressure amplifies the inter-sublattice hopping and leads to the degeneracy-lifted multipole fluctuations. Comprehensive investigations into these phenomena are still highly anticipated.

\begin{acknowledgments}
The authors are grateful to A.~Daido and S.~Sumita for fruitful discussions.
Some figures in this work were created by using {\sc FermiSurfer}~\cite{Kawamura2019fermisurfer}.
This work was supported by JSPS KAKENHI (Grants Nos. JP21K18145, 22H04933, JP22H01181, JP22KJ1716, JP23K17353).
\end{acknowledgments}

\bibliography{paper}

\clearpage

\renewcommand{\bibnumfmt}[1]{[S#1]}
\renewcommand{\citenumfont}[1]{S#1}
\renewcommand{\thesection}{S\arabic{section}}
\renewcommand{\theequation}{S\arabic{equation}}
\setcounter{equation}{0}
\renewcommand{\thefigure}{S\arabic{figure}}
\setcounter{figure}{0}
\renewcommand{\thetable}{S\arabic{table}}
\setcounter{table}{0}
\makeatletter
\c@secnumdepth = 2
\makeatother

\onecolumngrid

\begin{center}
 {\large \textmd{Supplemental Materials:} \\[0.3em]
 {\bfseries Field-induced superconductivity mediated by odd-parity multipole fluctuation}}
\end{center}

\setcounter{page}{1}

\section{Self-consistent condition for fluctuation exchange approximation}
The noninteracting Green functions for $U=0$ are expressed
by the $4\times4$ matrix form in the spin and sublattice basis,
\begin{equation}
  \label{eq:noint_green}
  G^{(0)}(\bm{k},i\omega_n) = \left(i\omega_n s_0\otimes\sigma_0 - \mathcal{H}(\bk)\right)^{-1},
\end{equation}
where $\omega_n=(2n+1)\pi T$ and $\mathcal{H}(\bk)$ are fermionic Matsubara frequencies and Hamiltonian.
Here, $s$ and $\sigma$ represent spin and sublattice degrees of freedom, respectively.
In the interacting case $U\neq0$, the dressed Green functions contain a self-energy $\Sigma(\bm{k},i\omega_n)$, 
\begin{align}
  \label{eq:green_function_dressed}
  G(\bm{k},i\omega_n) & = \left(i\omega_n s_0\otimes\sigma_0 - \mathcal{H}(\bk) -\Sigma(\bm{k},i\omega_n)\right)^{-1}.
\end{align}
In the FLEX approximation, the self-energy is expressed with the use of an effective interaction, $\Gamma^n(\bm{k},i\nu_n)$, as
\begin{align}
  \label{eq:self_energy}
  &\Sigma_{\xi\xi'}  (\bm{k},i\omega_n) 
                          = \frac{T}{N} \sum_{\bm{q},i\nu_n}
  \Gamma^n_{\xi\xi_1\xi'\xi_2}(\bm{q},i\nu_n)G_{\xi_1\xi_2}(\bm{k}-\bm{q},i\omega_n-i\nu_n),
\end{align}
and the effective interaction is given by
\begin{equation}
  \label{eq:effective_interaction}
  \Gamma^n_{\xi_1\xi_2\xi_3\xi_4}(\bm{k},i\nu_n) = U_{\xi_1\xi_2\xi_5\xi_6}\left(\chi_{\xi_5\xi_6\xi_7\xi_8}(\bm{k},i\nu_n)-\frac{1}{2}\chi^{(0)}_{\xi_5\xi_6\xi_7\xi_8}(\bm{k},i\nu_n)\right)U_{\xi_7\xi_8\xi_3\xi_4},
\end{equation}
where $U_{\xi_1\xi_2\xi_3\xi_4}$ is the bare interaction tensor which satisfies the following relation
\begin{align}
    \sum_{\xi_1\xi_2\xi_3\xi_4} U_{\xi_1\xi_2\xi_3\xi_4} c^\dagger_{\xi_1} c_{\xi_2} c_{\xi_3} c^\dagger_{\xi_4} = U \sum_{i,\sigma} n_{i\uparrow \sigma}n_{i\downarrow \sigma}, \\
  U_{\xi_1\xi_2\xi_3\xi_4} = \delta_{\sigma_1,\sigma_2}\delta_{\sigma_2,\sigma_3}\delta_{\sigma_3,\sigma_4} U_{s_1 s_2 s_3 s_4}, \\
  U_{\uparrow\downarrow\uparrow\downarrow} = U_{\downarrow\uparrow\downarrow\uparrow} = -U_{\uparrow\uparrow\downarrow\downarrow} = -U_{\downarrow\downarrow\uparrow\uparrow} = U,
\end{align}
and $i\nu_n$ are bosonic Matsubara frequencies.
Here, $\chi(\bm{k},i\nu_n)$ is the generalized susceptibility.
The abbreviated notation $\xi=(s,\sigma)$ are employed.
We introduce the bare susceptibility
\begin{align}
  \label{eq:bare_suscep}
  \chi^{(0)} _{\xi_1\xi_2\xi_3\xi_4}(\bm{q},i\nu_n)= -\frac{T}{N}\sum_{\bm{k},i\omega_n}G_{\xi_1\xi_3}(\bm{k},i\omega_n) G_{\xi_4\xi_2}(\bm{k}-\bm{q},i\omega_n-i\nu_n),
\end{align}
and compute the generalized susceptibility by
\begin{align}
  \label{eq:gener_suscep}
  \chi_{\xi_1\xi_2\xi_3\xi_4}(\bm{q},i\nu_n)=\chi^{(0)}_{\xi_1\xi_2\xi_3\xi_4}(\bm{q},i\nu_n)+\chi^{(0)}_{\xi_1\xi_2\xi_5\xi_6}(\bm{q},i\nu_n)U_{\xi_5\xi_6\xi_7\xi_8}\chi_{\xi_7\xi_8\xi_3\xi_4}(\bm{q},i\nu_n).
\end{align}
According to Eqs.~(\ref{eq:green_function_dressed})-(\ref{eq:gener_suscep}),
$G$, $\Sigma$, $\Gamma^n$, $\chi^{(0)}$, and $\chi$
depend on each other, and therefore, we self-consistently determine these functions. The FLEX approximation is a conserving approximation in which several conservation laws are satisfied in the framework of the Luttinger-Ward theory~\cite{Luttinger1960pr,Luttinger1960pr2,Baym1961pr,Baym1962pr}.

For functions with fermionic Matsubara frequencies $A(\bm{q},i \omega_n)$, the static limit 
$A(\bm{q},0)$ is evaluated by an approximation justified at low temperatures,
\begin{equation}
  A(\bm{q},0) \simeq \frac{A(\bm{q},i\pi T)+A(\bm{q},-i\pi T)}{2}.
\end{equation}

For the analysis of superconducting phase transition, particle-particle channel irreducible vertex function $\Gamma^a$ is needed, and it is 
obtained by
\begin{align}
  \Gamma^a_{\xi_1\xi_2\xi_3\xi_4}(\bm{q},i\nu_n) & = U_{\xi_1\xi_2\xi_3\xi_4}/2 +U_{\xi_1\xi_2\xi_5\xi_6}\chi_{\xi_5\xi_6\xi_7\xi_8}(\bm{q},i\nu_n)U_{\xi_7\xi_8\xi_3\xi_4}.
  \label{eq:pp_int}
\end{align}

\section{Multipole decomposition of susceptibility}
The normalized Pauli matrices ($\bar{\bm{\sigma}}=\bm{\sigma}/\sqrt{2}$) and the unit matrix ($\bar{\sigma}^0=\sigma^0/\sqrt{2}$) compose a complete basis of $2\times2$ matrix space.
Here, we adopt the normalization convention $\mathrm{tr}[(\sigma^{\mu})^\dagger\sigma^{\mu}]=1$.
Namely, any $2\times2$ complex matrix $A$ can be represented as a linear combination of the normalized Pauli matrices and the unit matrix,
\begin{equation}
    A = \sum_{\mu} a_{\mu} \bar{\sigma}^{\mu},
\end{equation}
where $a_{\mu}$ are complex coefficients.
When $A$ is a Hermitian matrix, $a_{\mu}$ should be real numbers.
The completeness of the Pauli matrices and the unit matrix leads to the following identity:
\begin{equation}
    \sum_{\mu} \bar{\sigma}^{\mu}_{ij} (\bar{\sigma}^{\mu}_{kl})^* = \sum_{\mu} \bar{\sigma}^{\mu}_{ij} \bar{\sigma}^{\mu}_{lk} = \delta_{ik}\delta_{jl},
\end{equation}
where $\delta_{ij}$ is Kronecker delta.
Here, the Hermite property of the Pauli and unit matrices $\bar{\sigma}_{ij} = (\bar{\sigma}_{ji})^*$ is used in the first equal sign.

The multipole operator in a two-sublattice system is represented by the tensor product of the Pauli matrices in spin- and sublattice-spaces: $\bar{\mathcal{Q}}^{\mu\nu}=\bar{s}^{\mu}\otimes\bar{\sigma}^{\nu}$.
Consequently, the following identity holds:
\begin{equation}
    \sum_{\mathcal{Q}} \bar{\mathcal{Q}}_{ij}\bar{\mathcal{Q}}_{kl} = \delta_{il}\delta_{jk}.
    \label{eq:multipole_identity}
\end{equation}
This identity facilitates the analysis of multipole-resolved fluctuations.
It should be noted that extending the current discussion to cases with general $N$ degrees of freedom is straightforward. This involves replacing the Pauli matrix with the $\mathfrak{su}(N)$ Lie algebra.

Upon inserting Eq.~(\ref{eq:multipole_identity}), Eq.~(\ref{eq:gener_suscep}) can be reformulated into a multipole-resolved form as,
\begin{align}
  \chi^{\mathcal{Q}}(\bm{q},i\nu_n) &= \bar{\mathcal{Q}}_{\xi_1\xi_2}\chi_{\xi_2\xi_1\xi_3\xi_4}(\bm{q},i\nu_n)\bar{\mathcal{Q}}_{\xi_3\xi_4} \notag \\
  &=\chi^{0, \mathcal{Q}}(\bm{q},i\nu_n) \notag \\
  &+\sum_{\mathcal{Q}'\mathcal{Q}''} \bar{\mathcal{Q}}_{\xi_1\xi_2}\chi^{0}_{\xi_2\xi_1\xi_5\xi_6}(\bm{q},i\nu_n) \bar{\mathcal{Q}}'_{\xi_5\xi_6}\bar{\mathcal{Q}}'_{\xi_7\xi_8}U_{\xi_8\xi_7\xi_9\xi_{10}}\bar{\mathcal{Q}}''_{\xi_{9}\xi_{10}}\bar{\mathcal{Q}}''_{\xi_{11}\xi_{12}}\chi_{\xi_{12}\xi_{11}\xi_3\xi_4}(\bm{q},i\nu_n)\bar{\mathcal{Q}}_{\xi_3\xi_4} \\
  &\approx\chi^{0, \mathcal{Q}}(\bm{q},i\nu_n) +\chi^{0, \mathcal{Q}}(\bm{q},i\nu_n)U^{\mathcal{Q}}\chi^{\mathcal{Q}}(\bm{q},i\nu_n),
  \label{eq:multipole_dyson}
\end{align}
where $U^{\mathcal{Q}} = \bar{\mathcal{Q}}_{\xi_1\xi_2}U_{\xi_1\xi_2\xi_3\xi_4}\bar{\mathcal{Q}}_{\xi_4\xi_3}$.
In this final expression, the cross terms between different multipole terms, denoted as $\chi^{\mathcal{Q}\mathcal{Q}'} = \bar{\mathcal{Q}}_{\xi_1\xi_2}\chi_{\xi_1\xi_2\xi_3\xi_4}\bar{\mathcal{Q}}'_{\xi_4\xi_3}$, are ignored.
Solving Eq.~(\ref{eq:multipole_dyson}), we obtain the enhanced multipole susceptibility due to interactions:
\begin{equation}
  \chi^{\mathcal{Q}}(q) \approx \frac{\chi^{0, \mathcal{Q}}(q)}{1-U^{\mathcal{Q}}\chi^{0, \mathcal{Q}}(q)}.
\end{equation}
A sufficient condition for achieving a large $\chi^{\mathcal{Q}}(q)$ entails having a large $\chi^{0, \mathcal{Q}}(q)$ and a positive $U^{\mathrm{Q}}$.
When we ignore the self-energy, the bare multipole susceptibility $\chi^{0, \mathcal{Q}}(q)$ can be expressed in the band basis as,
\begin{align}
    \chi^{0,\mathcal{Q}}(q) &= \bar{\mathcal{Q}}_{\beta\alpha}\chi^{(0)}_{\alpha\beta\gamma\delta}(q)\bar{\mathcal{Q}}_{\gamma\delta} \notag \\
    &= -\bar{\mathcal{Q}}_{\beta\alpha}\frac{T}{N}\sum_{k}G_{\alpha\gamma}(k)G_{\delta\beta}(k-q)\bar{\mathcal{Q}}_{\gamma\delta} \notag \\
    &= -\bar{\mathcal{Q}}_{\beta\alpha}\frac{T}{N}\sum_{k}U_{\alpha\zeta}(\bm{k})\mathcal{G}_{\zeta}(k)U^*_{\gamma\zeta}(\bm{k})U_{\delta\eta}(\bm{k}-\bm{q})\mathcal{G}_{\eta}(k-q)U^*_{\beta\eta}(\bm{k}-\bm{q})\bar{\mathcal{Q}}_{\gamma\delta} \notag \\
    &= -\frac{T}{N}\sum_{k}\left(U^{\dagger}(\bm{k}-\bm{q})\bar{\mathcal{Q}}U(\bm{k})\right)_{\eta\zeta}\mathcal{G}_{\zeta}(k)\mathcal{G}_{\eta}(k-q)\left(U^{\dagger}(\bm{k})\bar{\mathcal{Q}}U(\bm{k}-\bm{q})\right)_{\zeta\eta} \notag \\
    &=-\frac{1}{N} \sum_{\bk} \braket{u_{\eta,\bk-\bm{q}}|\bar{\mathcal{Q}}|u_{\zeta,\bk}} \braket{u_{\zeta,\bk}|\bar{\mathcal{Q}}|u_{\eta,\bk-\bm{q}}} \times\frac{f(\varepsilon_\eta(\bm{k}-\bm{q}))-f(\varepsilon_\zeta(\bm{k}))}{i\nu_n + \varepsilon_\eta(\bm{k}-\bm{q}) -\varepsilon_\zeta(\bm{k}) } \notag \\
    &=\sum_{\bk}\braket{u_{\eta,\bk-\bm{q}}|\bar{\mathcal{Q}}|u_{\zeta,\bk}} \braket{u_{\zeta,\bk}|\bar{\mathcal{Q}}|u_{\eta,\bk-\bm{q}}} L_{\zeta\eta}(\bm{k},\bm{q},i\nu_n),
\end{align}
where $L_{\zeta\eta}(\bm{k},\bm{q},i\nu_n)=-\frac{1}{N}\{f(\varepsilon_\eta(\bm{k}-\bm{q}))-f(\varepsilon_\zeta(\bm{k}))\}/\{i\nu_n + \varepsilon_\eta(\bm{k}-\bm{q}) -\varepsilon_\zeta(\bm{k})\}$ denotes the momentum-resolved Lindhard function between $\zeta$ and $\eta$ bands.
Here, $U(\bm{k})_{\alpha\zeta}=\braket{\alpha|u_{\zeta,\bm{k}}}$ represents the unitary matrix that diagonalizes Hamiltonian:
\begin{align}
    U^{\dagger}(\bm{k})\mathcal{H}(\bm{k})U(\bm{k}) = \mathcal{H}^{\mathrm{diag}}(\bm{k}), \\
    \mathcal{H}(\bm{k})\ket{u_{\zeta,\bm{k}}} = \varepsilon_{\zeta}(\bm{k})\ket{u_{\zeta,\bm{k}}}.
\end{align}
The Green function in the band basis is given by
\begin{align}
    \mathcal{G}(k) = U^\dagger(\bm{k})G(k)U(\bm{k}) \notag, \\
    \mathcal{G}_{\zeta}(k) = \frac{1}{i\omega_n-\varepsilon_\zeta(\bm{k})}.
\end{align}

In the same manner, we can decompose the effective interaction $\Gamma^n(q)$ [as defined in Eq.~(\ref{eq:effective_interaction})] and $\Gamma^a(q)$ [as defined in Eq.~(\ref{eq:pp_int})] into their respective multipole channels.
The decomposition is expressed as follows:
\begin{equation}
  \Gamma^{n,\mathcal{Q}}(q) \approx U^{\mathcal{Q}}\left(\chi^{\mathcal{Q}}(q)-\frac{1}{2}\chi^{0, \mathcal{Q}}(q)\right)U^{\mathcal{Q}},
  \label{eq:eff_int_n}
\end{equation}
\begin{equation}
  \Gamma^{a,\mathcal{Q}}(q) \approx \frac{U^{\mathcal{Q}}}{2} + U^{\mathcal{Q}}\chi^{\mathcal{Q}}(q)U^{\mathcal{Q}}.
  \label{eq:eff_int_a}
\end{equation}
In Eq.~(\ref{eq:eff_int_n}), the effective interaction for the particle-hole channel, $\Gamma^{n,\mathcal{Q}}(q)$, is expressed as a function of the multipole susceptibility $\chi^{\mathcal{Q}}(q)$ and the bare susceptibility $\chi^{0, \mathcal{Q}}(q)$ modulated by the interaction $U^{\mathcal{Q}}$.
Similarly, Eq.~(\ref{eq:eff_int_a}) depicts the effective interaction for the particle-particle channel, $\Gamma^{a,\mathcal{Q}}(q)$, also as a function of the multipole susceptibility and interaction.

\section{Cooper pairing channel from multipole fluctuations}
In this section, we give a comprehensive classification of Cooper pairing channels mediated by multipole fluctuations.
First, we summarize Cooper pairing channel decomposition in the presence of a single degree of freedom $\sigma$,
\begin{align}
    \mathcal{S}^{\sigma_0} &= \bar{\psi}_{\alpha} \sigma^0_{\alpha\beta} \psi_{\beta} V^{\sigma^0} \bar{\psi}_{\gamma} \sigma^0_{\gamma\delta} \psi_{\delta}=\frac{1}{2} V^{\sigma^0}\left\{\hat{\mathcal{P}}^{0,\dagger}\hat{\mathcal{P}}^{0} +\hat{\mathcal{P}}^{x,\dagger}\hat{\mathcal{P}}^{x} + \hat{\mathcal{P}}^{y,\dagger}\mathcal{P}^{y}+\hat{\mathcal{P}}^{z,\dagger}\mathcal{P}^{z}\right\}, \\
    \mathcal{S}^{\sigma_x} &= \bar{\psi}_{\alpha} \sigma^x_{\alpha\beta} \psi_{\beta} V^{\sigma^x} \bar{\psi}_{\gamma} \sigma^x_{\gamma\delta} \psi_{\delta}= \frac{1}{2}V^{\sigma^x}\left\{\hat{\mathcal{P}}^{0,\dagger}\hat{\mathcal{P}}^{0} +\hat{\mathcal{P}}^{x,\dagger}\hat{\mathcal{P}}^{x} - \hat{\mathcal{P}}^{y,\dagger}\hat{\mathcal{P}}^{y} -  \hat{\mathcal{P}}^{z,\dagger}\hat{\mathcal{P}}^{z}\right\}, \\
    \mathcal{S}^{\sigma_y} &= \bar{\psi}_{\alpha} \sigma^y_{\alpha\beta} \psi_{\beta} V^{\sigma^y} \bar{\psi}_{\gamma} \sigma^y_{\gamma\delta} \psi_{\delta}= \frac{1}{2}V^{\sigma^x}\left\{-\hat{\mathcal{P}}^{0,\dagger}\hat{\mathcal{P}}^{0} + \hat{\mathcal{P}}^{x,\dagger} \hat{\mathcal{P}}^{x} - \hat{\mathcal{P}}^{y,\dagger}\hat{\mathcal{P}}^{y} +  \hat{\mathcal{P}}^{z,\dagger}\hat{\mathcal{P}}^{z}\right\}, \\
    \mathcal{S}^{\sigma_z} &= \bar{\psi}_{\alpha} \sigma^z_{\alpha\beta} \psi_{\beta} V^{\sigma^z} \bar{\psi}_{\gamma} \sigma^z_{\gamma\delta} \psi_{\delta}= \frac{1}{2}V^{\sigma^x}\left\{\hat{\mathcal{P}}^{0,\dagger}\hat{\mathcal{P}}^{0} - \hat{\mathcal{P}}^{x,\dagger} \hat{\mathcal{P}}^{x}- \hat{\mathcal{P}}^{y,\dagger}\hat{\mathcal{P}}^{y} +  \hat{\mathcal{P}}^{z,\dagger}\hat{\mathcal{P}}^{z}\right\}, 
\end{align}
where the Cooper pair operators are defined by $\hat{\mathcal{P}}^{\mu}=\psi_{\alpha}\sigma^{\mu}_{\alpha\beta}\psi_{\beta}$.
Here, the following identities on the Pauli and unit matrix are used~\cite{Sumita2020prr}.
\begin{align}
    \sigma^0_{\alpha\beta}\sigma^0_{\gamma\delta} &= \frac{1}{2}\left(\sigma^0_{\alpha\gamma}\sigma^0_{\delta\beta}+\sigma^x_{\alpha\gamma}\sigma^x_{\delta\beta}+\sigma^y_{\alpha\gamma}\sigma^y_{\delta\beta}+\sigma^z_{\alpha\gamma}\sigma^z_{\delta\beta}\right), \\
    \sigma^x_{\alpha\beta}\sigma^x_{\gamma\delta} &= \frac{1}{2}\left(\sigma^0_{\alpha\gamma}\sigma^0_{\delta\beta}+\sigma^x_{\alpha\gamma}\sigma^x_{\delta\beta}-\sigma^y_{\alpha\gamma}\sigma^y_{\delta\beta}-\sigma^z_{\alpha\gamma}\sigma^z_{\delta\beta}\right), \\
    \sigma^y_{\alpha\beta}\sigma^y_{\gamma\delta} &= \frac{1}{2}\left(-\sigma^0_{\alpha\gamma}\sigma^0_{\delta\beta}+\sigma^x_{\alpha\gamma}\sigma^x_{\delta\beta}-\sigma^y_{\alpha\gamma}\sigma^y_{\delta\beta}+\sigma^z_{\alpha\gamma}\sigma^z_{\delta\beta}\right), \\
    \sigma^z_{\alpha\beta}\sigma^z_{\gamma\delta} &= \frac{1}{2}\left(\sigma^0_{\alpha\gamma}\sigma^0_{\delta\beta}-\sigma^x_{\alpha\gamma}\sigma^x_{\delta\beta}-\sigma^y_{\alpha\gamma}\sigma^y_{\delta\beta}+\sigma^z_{\alpha\gamma}\sigma^z_{\delta\beta}\right). 
\end{align}
Next, for the multipole composed of spin and sublattice degrees of freedom as in the case of our model, we obtain the Cooper pairing channel as follows:
\begin{align}
    \mathcal{S}^{\mathcal{Q}} &= \bar{\psi}_{\alpha} \mathcal{Q}^{\mu\nu}_{\alpha\beta} \psi_{\beta} V^{\mathcal{Q}} \bar{\psi}_{\gamma} \mathcal{Q}^{\mu\nu}_{\gamma\delta} \psi_{\delta} \\
    &= \bar{\psi}_{s_\alpha \sigma_\alpha} \bar{s}^{\mu}_{s_\alpha s_\beta}\bar{\sigma}^{\nu}_{\sigma_\alpha \sigma_\beta} \psi_{s_\beta \sigma_\beta} V^{\mathcal{Q}} \bar{\psi}_{s_\gamma \sigma_\gamma} \bar{s}^{\mu}_{s_\gamma s_\delta}\bar{\sigma}^{\nu}_{\sigma_\gamma \sigma_\delta} \psi_{s_\delta \sigma_\delta} \\
    &= V^{\mathcal{Q}} \bar{\psi}_{s_\alpha \sigma_\alpha}   \bar{\psi}_{s_\gamma \sigma_\gamma} \psi_{s_\delta \sigma_\delta} \psi_{s_\beta \sigma_\beta} \bar{s}^{\mu}_{s_\alpha s_\beta} \bar{s}^{\mu}_{s_\gamma s_\delta} \bar{\sigma}^{\nu}_{\sigma_\alpha \sigma_\beta} \bar{\sigma}^{\nu}_{\sigma_\gamma \sigma_\delta}.
\end{align}
For example, if we take $\mu=x$ and $\nu=y$, Cooper pairing channel is given as follows:
\begin{align}
     \mathcal{S}^{\mathcal{Q}^{xy}} 
    = \frac{V^{\mathcal{Q}}}{4} \big\{ 
    -&\hat{\mathcal{P}}^{00,\dagger} \hat{\mathcal{P}}^{00} 
    +\hat{\mathcal{P}}^{0x,\dagger} \hat{\mathcal{P}}^{0x}
    -\hat{\mathcal{P}}^{0y,\dagger} \hat{\mathcal{P}}^{0y}
    +\hat{\mathcal{P}}^{0z,\dagger} \hat{\mathcal{P}}^{0z}
    -\hat{\mathcal{P}}^{x0,\dagger} \hat{\mathcal{P}}^{x0}
    +\hat{\mathcal{P}}^{xx,\dagger} \hat{\mathcal{P}}^{xx}
    -\hat{\mathcal{P}}^{xy,\dagger} \hat{\mathcal{P}}^{xy}
    +\hat{\mathcal{P}}^{xz,\dagger} \hat{\mathcal{P}}^{xz} \notag \\
    +&\hat{\mathcal{P}}^{y0,\dagger} \hat{\mathcal{P}}^{y0}
    -\hat{\mathcal{P}}^{yx,\dagger} \hat{\mathcal{P}}^{yx}
    +\hat{\mathcal{P}}^{yy,\dagger} \hat{\mathcal{P}}^{yy}
    -\hat{\mathcal{P}}^{yz,\dagger} \hat{\mathcal{P}}^{yz}
    +\hat{\mathcal{P}}^{z0,\dagger} \hat{\mathcal{P}}^{z0}
    -\hat{\mathcal{P}}^{zx,\dagger} \hat{\mathcal{P}}^{zx}
    +\hat{\mathcal{P}}^{zy,\dagger} \hat{\mathcal{P}}^{zy}
    -\hat{\mathcal{P}}^{zz,\dagger} \hat{\mathcal{P}}^{zz} 
    \big\}.
\end{align}
Other decomposed Cooper pairing channels are summarized in Appendix.~\ref{sec:appendix}.

\section{\'{E}liashberg equation}
\begin{figure*}[t]
 \begin{center}
    \includegraphics[keepaspectratio, scale=0.25]{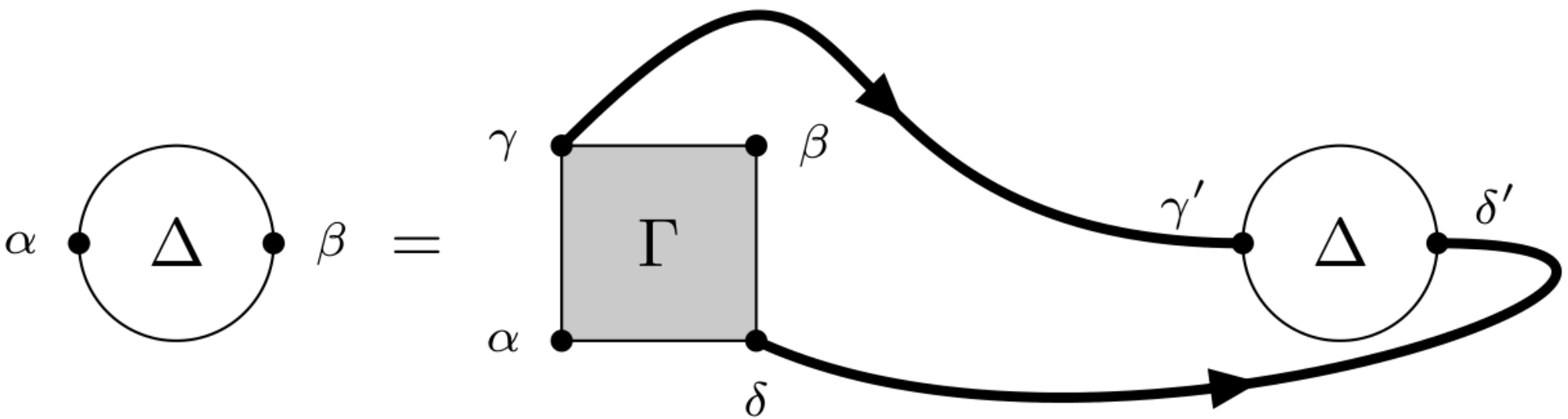}
  \end{center}
  \caption{The diagrammatic representation of the \'{E}liashberg equation.
  The shaded square and the black line represent the irreducible four-point vertex function in the particle-particle channel and the single-particle Green function, respectively.
  }
  \label{fig:eliash}
\end{figure*}
To investigate superconductivity, we adopt the linearized \'{E}liashberg equation which is expressed as
\begin{align}
  \lambda\Delta_{\alpha\beta}(k)  =
  \frac{T}{N}\sum_{k'} \Gamma^a_{\alpha \gamma \delta \beta}(k-k') G_{\gamma \gamma'}(k)\Delta_{\gamma' \delta'}(k) G_{\delta \delta'}(-k),
\end{align}
where $\Delta$ is the gap function and $\Gamma^a$ is the particle-particle channel irreducible vertex function obtained in Eq.~\eqref{eq:pp_int}. 
Figure~\ref{fig:eliash} shows the diagrammatic representation of the linearized \'{E}liashberg equation.
With the power method, we numerically evaluate $\lambda$, eigenvalues of the linearized \'{E}liashberg equation, and determine the critical temperature $T_{\rm c}$ from the criterion $\lambda=1$.

\section{Diagramatic consideration on the stability of odd-parity superconducting phase}
\begin{figure*}[t]
 \begin{center}
    \includegraphics[keepaspectratio, scale=0.5]{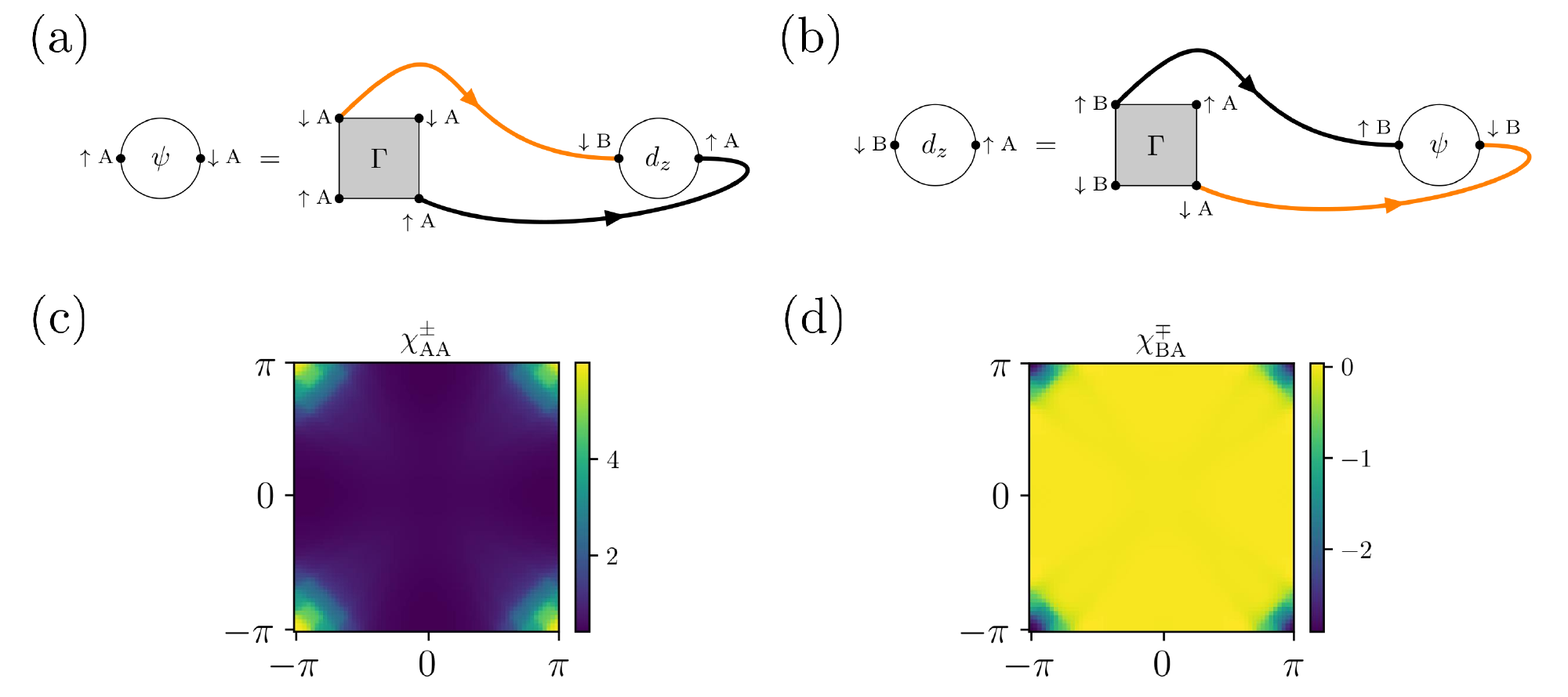}
  \end{center}
  \caption{
  (a) (b) The diagrammatic representation of the dominant scattering process between the Cooper pairs represented by $\psi^{\mathrm{AA}}$, $\psi^{\mathrm{BB}}$, and $d_z^{\mathrm{AB}}$.
  The black line and orange line represent the intra-sublattice Green function $G^{\mathrm{AA}}(k)$ and the inter-sublattice Green function $G^{\mathrm{AB}}(k)$, respectively.
  The scattering process between $d_z^{\mathrm{AB}}$  and $\psi^{\mathrm{AA}}$ or $\psi^{\mathrm{BB}}$ via the transverse magnetic fluctuation are shown.
  (c) The momentum dependence of the intra-sublattice transverse magnetic fluctuation which appears in the diagram (a).
  (d) The momentum dependence of the inter-sublattice transverse magnetic fluctuation which appears in the diagram (b).
  }
  \label{fig:diagram}
\end{figure*}
This section elucidates the distinctive role of inter-sublattice pairing in the field-induced odd-parity superconductivity. Utilizing the diagrammatic expression of the Éliashberg equation, we specify the important scattering process. While simplification is attained by solely considering the transverse spin fluctuation denoted by $\chi^{\pm}$ or $\chi^{\mp}$, expanding the following analysis to include longitudinal spin fluctuation $\chi^{zz}$ is straightforward.
In the following, we denote the dominant intra-sublattice spin-singlet component in the A and B sublattices as $\psi^{\mathrm{AA}}(\bm{k})$ and $\psi^{\mathrm{BB}}(\bm{k})$, respectively.

The scattering processes illustrated in Figs.~\ref{fig:diagram}(a-b) highlight how the unusual inter-sublattice pairing, represented by $\Im d_z^{\rm AB} (\bm{k})s_zis_y\otimes\sigma_y$, introduces the attractive force between $\psi^{\mathrm{AA}}(\bm{k})$ and $\psi^{\mathrm{BB}}(\bm{k})$. By amalgamating these two diagrams and tracing out the gap function $d^{\mathrm{BA}}_z$, a composite diagram elucidating the second-order scattering process between $\psi^{\mathrm{AA}}(\bm{k})$ and $\psi^{\mathrm{BB}}(\bm{k})$ is derived. 
Due to the positive sign of $\chi^{\pm}_{\mathrm{AA}}$ and the negative sign of $\chi^{\mp}_{\mathrm{BA}}$ [see Fig.~\ref{fig:diagram}(c,d)], the overall sign of this second-order scattering process is negative.
This scattering process with negative sign necessitates a sign change of gap functions through $2\bm{q}=(0,0)$ momentum transfer, a condition intrinsically met due to the relation $\psi^{\mathrm{AA}}(\bm{k})=-\psi^{\mathrm{BB}}(\bm{k})$. Notably, spin-orbit coupling is not required in this mechanism, thereby implying that field-induced superconductivity can be achieved in materials with weak spin-orbit coupling.

\appendix
\section{Decomposed Cooper channel for each augmented multipole fluctuation}
\label{sec:appendix}
\begin{align}
    \mathcal{S}^{\mathcal{Q}^{00}} 
    = \frac{V^{\mathcal{Q}}}{4} \big\{ 
    &\hat{\mathcal{P}}^{00,\dagger} \hat{\mathcal{P}}^{00} 
    +\hat{\mathcal{P}}^{0x,\dagger} \hat{\mathcal{P}}^{0x}
    +\hat{\mathcal{P}}^{0y,\dagger} \hat{\mathcal{P}}^{0y}
    +\hat{\mathcal{P}}^{0z,\dagger} \hat{\mathcal{P}}^{0z}
    +\hat{\mathcal{P}}^{x0,\dagger} \hat{\mathcal{P}}^{x0}
    +\hat{\mathcal{P}}^{xx,\dagger} \hat{\mathcal{P}}^{xx}
    +\hat{\mathcal{P}}^{xy,\dagger} \hat{\mathcal{P}}^{xy}
    +\hat{\mathcal{P}}^{xz,\dagger} \hat{\mathcal{P}}^{xz} \notag \\
    +&\hat{\mathcal{P}}^{y0,\dagger} \hat{\mathcal{P}}^{y0}
    +\hat{\mathcal{P}}^{yx,\dagger} \hat{\mathcal{P}}^{yx}
    +\hat{\mathcal{P}}^{yy,\dagger} \hat{\mathcal{P}}^{yy}
    +\hat{\mathcal{P}}^{yz,\dagger} \hat{\mathcal{P}}^{yz}
    +\hat{\mathcal{P}}^{z0,\dagger} \hat{\mathcal{P}}^{z0}
    +\hat{\mathcal{P}}^{zx,\dagger} \hat{\mathcal{P}}^{zx}
    +\hat{\mathcal{P}}^{zy,\dagger} \hat{\mathcal{P}}^{zy}
    +\hat{\mathcal{P}}^{zz,\dagger} \hat{\mathcal{P}}^{zz} 
    \big\} \\
    \mathcal{S}^{\mathcal{Q}^{0x}} 
    = \frac{V^{\mathcal{Q}}}{4} \big\{ 
    &\hat{\mathcal{P}}^{00,\dagger} \hat{\mathcal{P}}^{00} 
    +\hat{\mathcal{P}}^{0x,\dagger} \hat{\mathcal{P}}^{0x}
    -\hat{\mathcal{P}}^{0y,\dagger} \hat{\mathcal{P}}^{0y}
    -\hat{\mathcal{P}}^{0z,\dagger} \hat{\mathcal{P}}^{0z}
    +\hat{\mathcal{P}}^{x0,\dagger} \hat{\mathcal{P}}^{x0}
    +\hat{\mathcal{P}}^{xx,\dagger} \hat{\mathcal{P}}^{xx}
    -\hat{\mathcal{P}}^{xy,\dagger} \hat{\mathcal{P}}^{xy}
    -\hat{\mathcal{P}}^{xz,\dagger} \hat{\mathcal{P}}^{xz} \notag \\
    +&\hat{\mathcal{P}}^{y0,\dagger} \hat{\mathcal{P}}^{y0}
    +\hat{\mathcal{P}}^{yx,\dagger} \hat{\mathcal{P}}^{yx}
    -\hat{\mathcal{P}}^{yy,\dagger} \hat{\mathcal{P}}^{yy}
    -\hat{\mathcal{P}}^{yz,\dagger} \hat{\mathcal{P}}^{yz}
    +\hat{\mathcal{P}}^{z0,\dagger} \hat{\mathcal{P}}^{z0}
    +\hat{\mathcal{P}}^{zx,\dagger} \hat{\mathcal{P}}^{zx}
    -\hat{\mathcal{P}}^{zy,\dagger} \hat{\mathcal{P}}^{zy}
    -\hat{\mathcal{P}}^{zz,\dagger} \hat{\mathcal{P}}^{zz} 
    \big\} \\
    \mathcal{S}^{\mathcal{Q}^{0y}} 
    = \frac{V^{\mathcal{Q}}}{4} \big\{ 
    -&\hat{\mathcal{P}}^{00,\dagger} \hat{\mathcal{P}}^{00} 
    +\hat{\mathcal{P}}^{0x,\dagger} \hat{\mathcal{P}}^{0x}
    -\hat{\mathcal{P}}^{0y,\dagger} \hat{\mathcal{P}}^{0y}
    +\hat{\mathcal{P}}^{0z,\dagger} \hat{\mathcal{P}}^{0z}
    -\hat{\mathcal{P}}^{x0,\dagger} \hat{\mathcal{P}}^{x0}
    +\hat{\mathcal{P}}^{xx,\dagger} \hat{\mathcal{P}}^{xx}
    -\hat{\mathcal{P}}^{xy,\dagger} \hat{\mathcal{P}}^{xy}
    +\hat{\mathcal{P}}^{xz,\dagger} \hat{\mathcal{P}}^{xz} \notag \\
    -&\hat{\mathcal{P}}^{y0,\dagger} \hat{\mathcal{P}}^{y0}
    +\hat{\mathcal{P}}^{yx,\dagger} \hat{\mathcal{P}}^{yx}
    -\hat{\mathcal{P}}^{yy,\dagger} \hat{\mathcal{P}}^{yy}
    +\hat{\mathcal{P}}^{yz,\dagger} \hat{\mathcal{P}}^{yz}
    -\hat{\mathcal{P}}^{z0,\dagger} \hat{\mathcal{P}}^{z0}
    +\hat{\mathcal{P}}^{zx,\dagger} \hat{\mathcal{P}}^{zx}
    -\hat{\mathcal{P}}^{zy,\dagger} \hat{\mathcal{P}}^{zy}
    +\hat{\mathcal{P}}^{zz,\dagger} \hat{\mathcal{P}}^{zz} 
    \big\} \\
    \mathcal{S}^{\mathcal{Q}^{0z}} 
    = \frac{V^{\mathcal{Q}}}{4} \big\{ 
    &\hat{\mathcal{P}}^{00,\dagger} \hat{\mathcal{P}}^{00} 
    -\hat{\mathcal{P}}^{0x,\dagger} \hat{\mathcal{P}}^{0x}
    -\hat{\mathcal{P}}^{0y,\dagger} \hat{\mathcal{P}}^{0y}
    +\hat{\mathcal{P}}^{0z,\dagger} \hat{\mathcal{P}}^{0z}
    +\hat{\mathcal{P}}^{x0,\dagger} \hat{\mathcal{P}}^{x0}
    -\hat{\mathcal{P}}^{xx,\dagger} \hat{\mathcal{P}}^{xx}
    -\hat{\mathcal{P}}^{xy,\dagger} \hat{\mathcal{P}}^{xy}
    +\hat{\mathcal{P}}^{xz,\dagger} \hat{\mathcal{P}}^{xz} \notag \\
    +&\hat{\mathcal{P}}^{y0,\dagger} \hat{\mathcal{P}}^{y0}
    -\hat{\mathcal{P}}^{yx,\dagger} \hat{\mathcal{P}}^{yx}
    -\hat{\mathcal{P}}^{yy,\dagger} \hat{\mathcal{P}}^{yy}
    +\hat{\mathcal{P}}^{yz,\dagger} \hat{\mathcal{P}}^{yz}
    +\hat{\mathcal{P}}^{z0,\dagger} \hat{\mathcal{P}}^{z0}
    -\hat{\mathcal{P}}^{zx,\dagger} \hat{\mathcal{P}}^{zx}
    -\hat{\mathcal{P}}^{zy,\dagger} \hat{\mathcal{P}}^{zy}
    +\hat{\mathcal{P}}^{zz,\dagger} \hat{\mathcal{P}}^{zz} 
    \big\} \\
    \mathcal{S}^{\mathcal{Q}^{x0}} 
    = \frac{V^{\mathcal{Q}}}{4} \big\{ 
    &\hat{\mathcal{P}}^{00,\dagger} \hat{\mathcal{P}}^{00} 
    +\hat{\mathcal{P}}^{0x,\dagger} \hat{\mathcal{P}}^{0x}
    +\hat{\mathcal{P}}^{0y,\dagger} \hat{\mathcal{P}}^{0y}
    +\hat{\mathcal{P}}^{0z,\dagger} \hat{\mathcal{P}}^{0z}
    +\hat{\mathcal{P}}^{x0,\dagger} \hat{\mathcal{P}}^{x0}
    +\hat{\mathcal{P}}^{xx,\dagger} \hat{\mathcal{P}}^{xx}
    +\hat{\mathcal{P}}^{xy,\dagger} \hat{\mathcal{P}}^{xy}
    +\hat{\mathcal{P}}^{xz,\dagger} \hat{\mathcal{P}}^{xz} \notag \\
    -&\hat{\mathcal{P}}^{y0,\dagger} \hat{\mathcal{P}}^{y0}
    -\hat{\mathcal{P}}^{yx,\dagger} \hat{\mathcal{P}}^{yx}
    -\hat{\mathcal{P}}^{yy,\dagger} \hat{\mathcal{P}}^{yy}
    -\hat{\mathcal{P}}^{yz,\dagger} \hat{\mathcal{P}}^{yz}
    -\hat{\mathcal{P}}^{z0,\dagger} \hat{\mathcal{P}}^{z0}
    -\hat{\mathcal{P}}^{zx,\dagger} \hat{\mathcal{P}}^{zx}
    -\hat{\mathcal{P}}^{zy,\dagger} \hat{\mathcal{P}}^{zy}
    -\hat{\mathcal{P}}^{zz,\dagger} \hat{\mathcal{P}}^{zz} 
    \big\} \\
    \mathcal{S}^{\mathcal{Q}^{xx}} 
    = \frac{V^{\mathcal{Q}}}{4} \big\{ 
    &\hat{\mathcal{P}}^{00,\dagger} \hat{\mathcal{P}}^{00} 
    +\hat{\mathcal{P}}^{0x,\dagger} \hat{\mathcal{P}}^{0x}
    -\hat{\mathcal{P}}^{0y,\dagger} \hat{\mathcal{P}}^{0y}
    -\hat{\mathcal{P}}^{0z,\dagger} \hat{\mathcal{P}}^{0z}
    +\hat{\mathcal{P}}^{x0,\dagger} \hat{\mathcal{P}}^{x0}
    +\hat{\mathcal{P}}^{xx,\dagger} \hat{\mathcal{P}}^{xx}
    -\hat{\mathcal{P}}^{xy,\dagger} \hat{\mathcal{P}}^{xy}
    -\hat{\mathcal{P}}^{xz,\dagger} \hat{\mathcal{P}}^{xz} \notag \\
    -&\hat{\mathcal{P}}^{y0,\dagger} \hat{\mathcal{P}}^{y0}
    -\hat{\mathcal{P}}^{yx,\dagger} \hat{\mathcal{P}}^{yx}
    +\hat{\mathcal{P}}^{yy,\dagger} \hat{\mathcal{P}}^{yy}
    +\hat{\mathcal{P}}^{yz,\dagger} \hat{\mathcal{P}}^{yz}
    -\hat{\mathcal{P}}^{z0,\dagger} \hat{\mathcal{P}}^{z0}
    -\hat{\mathcal{P}}^{zx,\dagger} \hat{\mathcal{P}}^{zx}
    +\hat{\mathcal{P}}^{zy,\dagger} \hat{\mathcal{P}}^{zy}
    +\hat{\mathcal{P}}^{zz,\dagger} \hat{\mathcal{P}}^{zz} 
    \big\} \\
    \mathcal{S}^{\mathcal{Q}^{xy}} 
    = \frac{V^{\mathcal{Q}}}{4} \big\{ 
    -&\hat{\mathcal{P}}^{00,\dagger} \hat{\mathcal{P}}^{00} 
    +\hat{\mathcal{P}}^{0x,\dagger} \hat{\mathcal{P}}^{0x}
    -\hat{\mathcal{P}}^{0y,\dagger} \hat{\mathcal{P}}^{0y}
    +\hat{\mathcal{P}}^{0z,\dagger} \hat{\mathcal{P}}^{0z}
    -\hat{\mathcal{P}}^{x0,\dagger} \hat{\mathcal{P}}^{x0}
    +\hat{\mathcal{P}}^{xx,\dagger} \hat{\mathcal{P}}^{xx}
    -\hat{\mathcal{P}}^{xy,\dagger} \hat{\mathcal{P}}^{xy}
    +\hat{\mathcal{P}}^{xz,\dagger} \hat{\mathcal{P}}^{xz} \notag \\
    +&\hat{\mathcal{P}}^{y0,\dagger} \hat{\mathcal{P}}^{y0}
    -\hat{\mathcal{P}}^{yx,\dagger} \hat{\mathcal{P}}^{yx}
    +\hat{\mathcal{P}}^{yy,\dagger} \hat{\mathcal{P}}^{yy}
    -\hat{\mathcal{P}}^{yz,\dagger} \hat{\mathcal{P}}^{yz}
    +\hat{\mathcal{P}}^{z0,\dagger} \hat{\mathcal{P}}^{z0}
    -\hat{\mathcal{P}}^{zx,\dagger} \hat{\mathcal{P}}^{zx}
    +\hat{\mathcal{P}}^{zy,\dagger} \hat{\mathcal{P}}^{zy}
    -\hat{\mathcal{P}}^{zz,\dagger} \hat{\mathcal{P}}^{zz} 
    \big\} \\
    \mathcal{S}^{\mathcal{Q}^{xz}} 
    = \frac{V^{\mathcal{Q}}}{4} \big\{ 
    &\hat{\mathcal{P}}^{00,\dagger} \hat{\mathcal{P}}^{00} 
    -\hat{\mathcal{P}}^{0x,\dagger} \hat{\mathcal{P}}^{0x}
    -\hat{\mathcal{P}}^{0y,\dagger} \hat{\mathcal{P}}^{0y}
    +\hat{\mathcal{P}}^{0z,\dagger} \hat{\mathcal{P}}^{0z}
    +\hat{\mathcal{P}}^{x0,\dagger} \hat{\mathcal{P}}^{x0}
    -\hat{\mathcal{P}}^{xx,\dagger} \hat{\mathcal{P}}^{xx}
    -\hat{\mathcal{P}}^{xy,\dagger} \hat{\mathcal{P}}^{xy}
    +\hat{\mathcal{P}}^{xz,\dagger} \hat{\mathcal{P}}^{xz} \notag \\
    -&\hat{\mathcal{P}}^{y0,\dagger} \hat{\mathcal{P}}^{y0}
    +\hat{\mathcal{P}}^{yx,\dagger} \hat{\mathcal{P}}^{yx}
    +\hat{\mathcal{P}}^{yy,\dagger} \hat{\mathcal{P}}^{yy}
    -\hat{\mathcal{P}}^{yz,\dagger} \hat{\mathcal{P}}^{yz}
    -\hat{\mathcal{P}}^{z0,\dagger} \hat{\mathcal{P}}^{z0}
    +\hat{\mathcal{P}}^{zx,\dagger} \hat{\mathcal{P}}^{zx}
    +\hat{\mathcal{P}}^{zy,\dagger} \hat{\mathcal{P}}^{zy}
    -\hat{\mathcal{P}}^{zz,\dagger} \hat{\mathcal{P}}^{zz} 
    \big\} \\
    \mathcal{S}^{\mathcal{Q}^{y0}} 
    = \frac{V^{\mathcal{Q}}}{4} \big\{ 
    -&\hat{\mathcal{P}}^{00,\dagger} \hat{\mathcal{P}}^{00} 
    -\hat{\mathcal{P}}^{0x,\dagger} \hat{\mathcal{P}}^{0x}
    -\hat{\mathcal{P}}^{0y,\dagger} \hat{\mathcal{P}}^{0y}
    -\hat{\mathcal{P}}^{0z,\dagger} \hat{\mathcal{P}}^{0z}
    +\hat{\mathcal{P}}^{x0,\dagger} \hat{\mathcal{P}}^{x0}
    +\hat{\mathcal{P}}^{xx,\dagger} \hat{\mathcal{P}}^{xx}
    +\hat{\mathcal{P}}^{xy,\dagger} \hat{\mathcal{P}}^{xy}
    +\hat{\mathcal{P}}^{xz,\dagger} \hat{\mathcal{P}}^{xz} \notag \\
    -&\hat{\mathcal{P}}^{y0,\dagger} \hat{\mathcal{P}}^{y0}
    -\hat{\mathcal{P}}^{yx,\dagger} \hat{\mathcal{P}}^{yx}
    -\hat{\mathcal{P}}^{yy,\dagger} \hat{\mathcal{P}}^{yy}
    -\hat{\mathcal{P}}^{yz,\dagger} \hat{\mathcal{P}}^{yz}
    +\hat{\mathcal{P}}^{z0,\dagger} \hat{\mathcal{P}}^{z0}
    +\hat{\mathcal{P}}^{zx,\dagger} \hat{\mathcal{P}}^{zx}
    +\hat{\mathcal{P}}^{zy,\dagger} \hat{\mathcal{P}}^{zy}
    +\hat{\mathcal{P}}^{zz,\dagger} \hat{\mathcal{P}}^{zz} 
    \big\} \\
    \mathcal{S}^{\mathcal{Q}^{yx}} 
    = \frac{V^{\mathcal{Q}}}{4} \big\{ 
    -&\hat{\mathcal{P}}^{00,\dagger} \hat{\mathcal{P}}^{00} 
    -\hat{\mathcal{P}}^{0x,\dagger} \hat{\mathcal{P}}^{0x}
    +\hat{\mathcal{P}}^{0y,\dagger} \hat{\mathcal{P}}^{0y}
    +\hat{\mathcal{P}}^{0z,\dagger} \hat{\mathcal{P}}^{0z}
    +\hat{\mathcal{P}}^{x0,\dagger} \hat{\mathcal{P}}^{x0}
    +\hat{\mathcal{P}}^{xx,\dagger} \hat{\mathcal{P}}^{xx}
    -\hat{\mathcal{P}}^{xy,\dagger} \hat{\mathcal{P}}^{xy}
    -\hat{\mathcal{P}}^{xz,\dagger} \hat{\mathcal{P}}^{xz} \notag \\
    -&\hat{\mathcal{P}}^{y0,\dagger} \hat{\mathcal{P}}^{y0}
    -\hat{\mathcal{P}}^{yx,\dagger} \hat{\mathcal{P}}^{yx}
    +\hat{\mathcal{P}}^{yy,\dagger} \hat{\mathcal{P}}^{yy}
    +\hat{\mathcal{P}}^{yz,\dagger} \hat{\mathcal{P}}^{yz}
    +\hat{\mathcal{P}}^{z0,\dagger} \hat{\mathcal{P}}^{z0}
    +\hat{\mathcal{P}}^{zx,\dagger} \hat{\mathcal{P}}^{zx}
    -\hat{\mathcal{P}}^{zy,\dagger} \hat{\mathcal{P}}^{zy}
    -\hat{\mathcal{P}}^{zz,\dagger} \hat{\mathcal{P}}^{zz} 
    \big\} \\
    \mathcal{S}^{\mathcal{Q}^{yy}} 
    = \frac{V^{\mathcal{Q}}}{4} \big\{ 
    &\hat{\mathcal{P}}^{00,\dagger} \hat{\mathcal{P}}^{00} 
    -\hat{\mathcal{P}}^{0x,\dagger} \hat{\mathcal{P}}^{0x}
    +\hat{\mathcal{P}}^{0y,\dagger} \hat{\mathcal{P}}^{0y}
    -\hat{\mathcal{P}}^{0z,\dagger} \hat{\mathcal{P}}^{0z}
    -\hat{\mathcal{P}}^{x0,\dagger} \hat{\mathcal{P}}^{x0}
    +\hat{\mathcal{P}}^{xx,\dagger} \hat{\mathcal{P}}^{xx}
    -\hat{\mathcal{P}}^{xy,\dagger} \hat{\mathcal{P}}^{xy}
    +\hat{\mathcal{P}}^{xz,\dagger} \hat{\mathcal{P}}^{xz} \notag \\
    +&\hat{\mathcal{P}}^{y0,\dagger} \hat{\mathcal{P}}^{y0}
    -\hat{\mathcal{P}}^{yx,\dagger} \hat{\mathcal{P}}^{yx}
    +\hat{\mathcal{P}}^{yy,\dagger} \hat{\mathcal{P}}^{yy}
    -\hat{\mathcal{P}}^{yz,\dagger} \hat{\mathcal{P}}^{yz}
    -\hat{\mathcal{P}}^{z0,\dagger} \hat{\mathcal{P}}^{z0}
    +\hat{\mathcal{P}}^{zx,\dagger} \hat{\mathcal{P}}^{zx}
    -\hat{\mathcal{P}}^{zy,\dagger} \hat{\mathcal{P}}^{zy}
    +\hat{\mathcal{P}}^{zz,\dagger} \hat{\mathcal{P}}^{zz} 
    \big\} \\
    \mathcal{S}^{\mathcal{Q}^{yz}} 
    = \frac{V^{\mathcal{Q}}}{4} \big\{ 
    -&\hat{\mathcal{P}}^{00,\dagger} \hat{\mathcal{P}}^{00} 
    +\hat{\mathcal{P}}^{0x,\dagger} \hat{\mathcal{P}}^{0x}
    +\hat{\mathcal{P}}^{0y,\dagger} \hat{\mathcal{P}}^{0y}
    -\hat{\mathcal{P}}^{0z,\dagger} \hat{\mathcal{P}}^{0z}
    +\hat{\mathcal{P}}^{x0,\dagger} \hat{\mathcal{P}}^{x0}
    -\hat{\mathcal{P}}^{xx,\dagger} \hat{\mathcal{P}}^{xx}
    -\hat{\mathcal{P}}^{xy,\dagger} \hat{\mathcal{P}}^{xy}
    +\hat{\mathcal{P}}^{xz,\dagger} \hat{\mathcal{P}}^{xz} \notag \\
    -&\hat{\mathcal{P}}^{y0,\dagger} \hat{\mathcal{P}}^{y0}
    +\hat{\mathcal{P}}^{yx,\dagger} \hat{\mathcal{P}}^{yx}
    +\hat{\mathcal{P}}^{yy,\dagger} \hat{\mathcal{P}}^{yy}
    -\hat{\mathcal{P}}^{yz,\dagger} \hat{\mathcal{P}}^{yz}
    +\hat{\mathcal{P}}^{z0,\dagger} \hat{\mathcal{P}}^{z0}
    -\hat{\mathcal{P}}^{zx,\dagger} \hat{\mathcal{P}}^{zx}
    -\hat{\mathcal{P}}^{zy,\dagger} \hat{\mathcal{P}}^{zy}
    +\hat{\mathcal{P}}^{zz,\dagger} \hat{\mathcal{P}}^{zz} 
    \big\} \\
    \mathcal{S}^{\mathcal{Q}^{z0}} 
    = \frac{V^{\mathcal{Q}}}{4} \big\{ 
    &\hat{\mathcal{P}}^{00,\dagger} \hat{\mathcal{P}}^{00} 
    +\hat{\mathcal{P}}^{0x,\dagger} \hat{\mathcal{P}}^{0x}
    +\hat{\mathcal{P}}^{0y,\dagger} \hat{\mathcal{P}}^{0y}
    +\hat{\mathcal{P}}^{0z,\dagger} \hat{\mathcal{P}}^{0z}
    -\hat{\mathcal{P}}^{x0,\dagger} \hat{\mathcal{P}}^{x0}
    -\hat{\mathcal{P}}^{xx,\dagger} \hat{\mathcal{P}}^{xx}
    -\hat{\mathcal{P}}^{xy,\dagger} \hat{\mathcal{P}}^{xy}
    -\hat{\mathcal{P}}^{xz,\dagger} \hat{\mathcal{P}}^{xz} \notag \\
    -&\hat{\mathcal{P}}^{y0,\dagger} \hat{\mathcal{P}}^{y0}
    -\hat{\mathcal{P}}^{yx,\dagger} \hat{\mathcal{P}}^{yx}
    -\hat{\mathcal{P}}^{yy,\dagger} \hat{\mathcal{P}}^{yy}
    -\hat{\mathcal{P}}^{yz,\dagger} \hat{\mathcal{P}}^{yz}
    +\hat{\mathcal{P}}^{z0,\dagger} \hat{\mathcal{P}}^{z0}
    +\hat{\mathcal{P}}^{zx,\dagger} \hat{\mathcal{P}}^{zx}
    +\hat{\mathcal{P}}^{zy,\dagger} \hat{\mathcal{P}}^{zy}
    +\hat{\mathcal{P}}^{zz,\dagger} \hat{\mathcal{P}}^{zz} 
    \big\} \\
    \mathcal{S}^{\mathcal{Q}^{zx}} 
    = \frac{V^{\mathcal{Q}}}{4} \big\{ 
    &\hat{\mathcal{P}}^{00,\dagger} \hat{\mathcal{P}}^{00} 
    +\hat{\mathcal{P}}^{0x,\dagger} \hat{\mathcal{P}}^{0x}
    -\hat{\mathcal{P}}^{0y,\dagger} \hat{\mathcal{P}}^{0y}
    -\hat{\mathcal{P}}^{0z,\dagger} \hat{\mathcal{P}}^{0z}
    -\hat{\mathcal{P}}^{x0,\dagger} \hat{\mathcal{P}}^{x0}
    -\hat{\mathcal{P}}^{xx,\dagger} \hat{\mathcal{P}}^{xx}
    +\hat{\mathcal{P}}^{xy,\dagger} \hat{\mathcal{P}}^{xy}
    +\hat{\mathcal{P}}^{xz,\dagger} \hat{\mathcal{P}}^{xz} \notag \\
    -&\hat{\mathcal{P}}^{y0,\dagger} \hat{\mathcal{P}}^{y0}
    -\hat{\mathcal{P}}^{yx,\dagger} \hat{\mathcal{P}}^{yx}
    +\hat{\mathcal{P}}^{yy,\dagger} \hat{\mathcal{P}}^{yy}
    +\hat{\mathcal{P}}^{yz,\dagger} \hat{\mathcal{P}}^{yz}
    +\hat{\mathcal{P}}^{z0,\dagger} \hat{\mathcal{P}}^{z0}
    +\hat{\mathcal{P}}^{zx,\dagger} \hat{\mathcal{P}}^{zx}
    -\hat{\mathcal{P}}^{zy,\dagger} \hat{\mathcal{P}}^{zy}
    -\hat{\mathcal{P}}^{zz,\dagger} \hat{\mathcal{P}}^{zz} 
    \big\} \\
    \mathcal{S}^{\mathcal{Q}^{zy}} 
    = \frac{V^{\mathcal{Q}}}{4} \big\{ 
    -&\hat{\mathcal{P}}^{00,\dagger} \hat{\mathcal{P}}^{00} 
    +\hat{\mathcal{P}}^{0x,\dagger} \hat{\mathcal{P}}^{0x}
    -\hat{\mathcal{P}}^{0y,\dagger} \hat{\mathcal{P}}^{0y}
    +\hat{\mathcal{P}}^{0z,\dagger} \hat{\mathcal{P}}^{0z}
    +\hat{\mathcal{P}}^{x0,\dagger} \hat{\mathcal{P}}^{x0}
    -\hat{\mathcal{P}}^{xx,\dagger} \hat{\mathcal{P}}^{xx}
    +\hat{\mathcal{P}}^{xy,\dagger} \hat{\mathcal{P}}^{xy}
    -\hat{\mathcal{P}}^{xz,\dagger} \hat{\mathcal{P}}^{xz} \notag \\
    +&\hat{\mathcal{P}}^{y0,\dagger} \hat{\mathcal{P}}^{y0}
    -\hat{\mathcal{P}}^{yx,\dagger} \hat{\mathcal{P}}^{yx}
    +\hat{\mathcal{P}}^{yy,\dagger} \hat{\mathcal{P}}^{yy}
    -\hat{\mathcal{P}}^{yz,\dagger} \hat{\mathcal{P}}^{yz}
    -\hat{\mathcal{P}}^{z0,\dagger} \hat{\mathcal{P}}^{z0}
    +\hat{\mathcal{P}}^{zx,\dagger} \hat{\mathcal{P}}^{zx}
    -\hat{\mathcal{P}}^{zy,\dagger} \hat{\mathcal{P}}^{zy}
    +\hat{\mathcal{P}}^{zz,\dagger} \hat{\mathcal{P}}^{zz} 
    \big\} \\
    \mathcal{S}^{\mathcal{Q}^{0z}} 
    = \frac{V^{\mathcal{Q}}}{4} \big\{ 
    &\hat{\mathcal{P}}^{00,\dagger} \hat{\mathcal{P}}^{00} 
    -\hat{\mathcal{P}}^{0x,\dagger} \hat{\mathcal{P}}^{0x}
    -\hat{\mathcal{P}}^{0y,\dagger} \hat{\mathcal{P}}^{0y}
    +\hat{\mathcal{P}}^{0z,\dagger} \hat{\mathcal{P}}^{0z}
    +\hat{\mathcal{P}}^{x0,\dagger} \hat{\mathcal{P}}^{x0}
    -\hat{\mathcal{P}}^{xx,\dagger} \hat{\mathcal{P}}^{xx}
    -\hat{\mathcal{P}}^{xy,\dagger} \hat{\mathcal{P}}^{xy}
    +\hat{\mathcal{P}}^{xz,\dagger} \hat{\mathcal{P}}^{xz} \notag \\
    +&\hat{\mathcal{P}}^{y0,\dagger} \hat{\mathcal{P}}^{y0}
    -\hat{\mathcal{P}}^{yx,\dagger} \hat{\mathcal{P}}^{yx}
    -\hat{\mathcal{P}}^{yy,\dagger} \hat{\mathcal{P}}^{yy}
    +\hat{\mathcal{P}}^{yz,\dagger} \hat{\mathcal{P}}^{yz}
    +\hat{\mathcal{P}}^{z0,\dagger} \hat{\mathcal{P}}^{z0}
    -\hat{\mathcal{P}}^{zx,\dagger} \hat{\mathcal{P}}^{zx}
    -\hat{\mathcal{P}}^{zy,\dagger} \hat{\mathcal{P}}^{zy}
    +\hat{\mathcal{P}}^{zz,\dagger} \hat{\mathcal{P}}^{zz} 
    \big\} 
\end{align}

\end{document}